\documentclass[prd,aps,a4paper,eqsecnum,floatfix,nofootinbib,twocolumn]{revtex4}  

\usepackage{graphicx} 
\usepackage{amsmath,amsfonts,amssymb}
\usepackage{placeins}
\usepackage{appendix}
\usepackage[caption=false,justification=justified]{subfig}

\newcommand{\beq}{\begin{equation}}
\newcommand{\eeq}{\end{equation}}
\newcommand{\bea}{\begin{eqnarray}}
\newcommand{\eea}{\end{eqnarray}}
\newcommand{\be}{\begin{equation}}      
\newcommand{\ee}{\end{equation}}
\def\nn{\nonumber}

\begin{document}

\title{5d Schwarzschild-Tangherlini spacetime: MST-like  formalism for a Reduced Confluent Heun Equation}

\author{Donato Bini$^{1}$, Giorgio Di Russo$^{2}$, Veronica Fantini$^3$}   
  \affiliation{
$^1$Istituto per le Applicazioni del Calcolo ``M. Picone,'' CNR, I-00185 Rome, Italy\\
$^2$School of Fundamental Physics and Mathematical Sciences, Hangzhou Institute for Advanced Study, UCAS, Hangzhou 310024, China\\
$^3$Universit\'e Paris-Saclay, CNRS, Laboratoire de Mathematique d'Orsay (LMO), 91405 Orsay, France
}

\date{\today}

\begin{abstract}
We study the five-dimensional Schwarzschild-Tangherlini solution, with particular attention to its geodesic structure and massless scalar perturbations. In the probe limit, we present two applications. First, we compute the scattering angle for unbound geodesics showing both  post-Newtonian and post-Minkowskian type expansions, and succeeding in resumming the resulting series in terms of hypergeometric functions. 
Second, we derive the Lyapunov exponent for deviations from a critical circular orbit, which is relevant to the eikonal estimation of quasinormal modes.

We then investigate the dynamics of massless scalar $(s=0)$ perturbations, for which the radial equation becomes a Reduced Confluent Heun equation. In this $d=5$ Schwarzschild-Tangherlini case we develop an original extension of the standard Mano-Suzuki-Takasugi (MST) formalism and validate the construction by computing the renormalized angular-momentum parameter $\nu$, whose value agrees with an independent determination based on the quantum Seiberg-Witten formalism. Finally, we analyze the energy flux from circular orbits, obtaining post-Newtonian results through 2.5PN order.

\end{abstract}

\maketitle 

\section{Introduction}

The study of gravity in spacetime dimensions greater than four has become an important area of research in theoretical physics, motivated by developments in string theory, brane-world scenarios, and attempts to formulate a consistent quantum theory of gravity. Higher-dimensional gravitational theories often exhibit novel phenomena absent in four dimensions, including new black hole topologies, modified stability properties, and richer solution spaces. 

Among the most fundamental vacuum solutions is the Schwarzschild-Tangherlini (ST) metric, which generalizes the four-dimensional Schwarzschild solution to an arbitrary spacetime dimension $d$ \cite{Tangherlini:1963bw}. The ST solution describes a static, spherically symmetric black hole in a $d$-dimensional asymptotically flat spacetime, and constitutes the simplest black hole solution of the vacuum Einstein equations in higher dimensions. It serves as the starting point for the construction and analysis of more complicated geometries, including rotating black holes, charged solutions, black strings, black branes, etc. \cite{HOROWITZ1991197,Gregory:1993vy}.

In five spacetime dimensions the Schwarzschild-Tangherlini metric (STd5) acquires several features that distinguish it from its four-dimensional counterpart. The event horizon possesses the topology of a three-sphere $S^3$, and the gravitational potential (deduced from the asymptotic properties of the metric) exhibits a different radial dependence, reflecting the altered structure of Newtonian gravity in higher dimensions. Furthermore, five-dimensional gravity provides a particularly interesting setting because it is the lowest dimension in which genuinely new black hole solutions, such as the Myers-Perry black holes  \cite{Myers:1986un} and black rings, exist. Consequently, the STd5 solution serves as a natural reference geometry for exploring the broader landscape of higher-dimensional black hole physics.

The purpose of this paper is to investigate, from both a geometrical and a physical perspective, the five-dimensional Schwarzschild-Tangherlini spacetime, illustrating both the similarities and the essential differences between four-dimensional and higher-dimensional gravity.

We begin by reviewing the basic geometrical properties of the spacetime and establishing the notation used throughout the paper.

The paper contains several new results already at the point-particle (probe) level. In particular, we derive an explicit expression for the scattering angle along hyperbolic-like trajectories, presenting both a Post-Newtonian (PN) type expansion and a Post-Minkowskian (PM) type of resummations. We check and confirm the expression of the scattering angle by a suitable, closed form resummation formula starting from  the radial action  for null geodesics, generalizing to the present situation results valid in 4d.  We  study  geodesic deviations form a fiducial \lq\lq reference" circular geodesic (for simplicity), discussing comparisons with the standard 4d Schwarzschild spacetime case. Finally, we provide an eikonal estimation of the quasinormal modes (QNM) spectrum.

The main focus of the paper, however, is the study of massless scalar perturbations of the five-dimensional background.

We consider the massless scalar field equation both in the source-free case and in the presence of a scalar point-particle source moving on a circular orbit. The field equations are separated with respect to time (associated with a timelike Killing vector), two azimuthal coordinates (associated with two spatial Killing vectors generating closed orbits), one polar coordinate, and the radial coordinate.

Owing to the symmetries of the background, the angular sector is naturally described by a generalization of the spherical harmonics on $S^3$, namely the Wigner rotation matrices. It is the analysis of the radial equation, however, which constitutes the core of the present work.

Unlike the familiar four-dimensional Schwarzschild case, where the radial equation reduces to a confluent Heun equation (CHE), the corresponding equation in the five-dimensional Schwarzschild-Tangherlini spacetime is a reduced confluent Heun equation (RCHE). It is worth to recall that this differential equation (i.e., the RCHE) is known to be useful in various applications; for example, it describes radial and angular equations of some fuzzball solutions  such as JMaRT and D1-D5 circular string profile fuzzballs \cite{Cardoso:2005gj,Bianchi:2022qph,Bianchi:2019lmi,Bianchi:2023rlt,DiRusso:2024hmd}.

Similarly, the CHE arises not only in the study of scalar perturbations of black holes in the Kerr-Newman family, but also in the analysis of the so-called Topological Stars and W solitons, both of which are five-dimensional solutions of the Einstein-Maxwell equations. See Refs. \cite{Bah:2020pdz,Bianchi:2023sfs,Heidmann:2023ojf,Cipriani:2024ygw,Bianchi:2024vmi,Bianchi:2024rod,DiRusso:2025lip,Heidmann:2025pbb,Bianchi:2025uis,Dima:2025tjz} for recent developments.

As is well known, taking confluent limits requires particular care because they involve the coalescence of singular points. Consequently, one cannot simply solve the ordinary confluent Heun equation as in the four-dimensional Schwarzschild case, where the Mano-Suzuki-Takasugi (MST) formalism \cite{Mano:1996mf,Mano:1996vt,Sasaki:2003xr} applies, and subsequently perform a \lq\lq confluence procedure." Such an approach would amount to treating only a simplified limiting problem and, in general, does not reproduce the correct RCHE. Instead, the MST formalism must be generalized directly at the level of the RCHE (one of the goals of the present research).

The first main result of the present paper is precisely the construction of such a generalized MST formalism. As a non-trivial consistency check, we compare the renormalized angular momentum parameter $\nu$ obtained within our generalized MST approach with the corresponding quantity computed using the quantum Seiberg-Witten (qSW or simply SW) formalism. The complete agreement between the two determinations of $\nu$ provides a strong evidence for the correctness of the proposed formalism.

As a concrete application, we compute the scalar energy flux emitted by a point particle moving on a circular orbit within a newly developed five-dimensional gravitational self-force framework (working, as usual, in PN sense). Explicit Post-Newtonian results are presented for the monopole, dipole, quadrupole and octupole modes. Although the present work is restricted to relatively low PN orders (i.e., $O(\eta^{10})$, where $\eta=\frac{1}{c}$ denotes a place-holder for PN expansion), the formalism developed here can be systematically extended to arbitrarily high PN accuracy.

Unless otherwise stated, we use units such that $c=1$ and $G_{d}=1$ (unless differently specified). Our chosen metric signature is the mostly positive one $(-+++)$.

\section{Metric}

We write the five-dimensional Schwarzschild-Tangherlini metric in the form
\beq
ds^2={}_{(2)} ds_{(t,r)}^2+{}_{(3)} ds_{(\theta,\phi,\psi)}^2\,,
\eeq
where
\bea
\label{met_split}
{}_{(2)} ds_{(t,r)}^2&=&-\left(1-\frac{M^2}{r^2} \right)dt^2+\frac{dr^2}{\left(1-\frac{M^2}{r^2} \right)}\,,\nonumber\\
{}_{(3)} ds_{(\theta,\phi,\psi)}^2&=& r^2(d\theta^2 +\sin^2\theta d\phi^2+\cos^2\theta d\psi^2)\,,\qquad
\eea
and the range of variability of the coordinates is the following:
$t\in {\mathbb R}$, $r>M$, $0<\theta <\pi/2$, $\phi\in [0,2\pi)$, $\psi\in [0,2\pi)$.
The limiting cases $r=M$, $\theta=\pi/2$ (especially relevant because allows to cast the metric in a form close to four dimensional Schwarzschild metric), etc. are still allowed but require some care, as we will discuss  in detail  below.

Furthermore, for a later use, let us introduce the notation
\beq
f(r)= 1-\frac{M^2}{r^2}\,,\quad [s,c]=[\sin \theta, \cos\theta] \,.
\eeq
The STd5 spacetime is Ricci-flat but the Riemann tensor is nonzero and the only non-vanishing components (modulo symmetries) are given by
\bea
\label{compo}
{}[R_{t\phi t\phi},R_{t \psi t \psi},R_{t\theta t \theta},R_{trtr}]  &=& -f \frac{M^2}{r^2}\big[s^2,-c^2, \frac1{f^2},-\frac{3}{r^2f}\Big]\,,\nonumber\\
{}[R_{r\phi r \phi}, R_{r\psi r \psi},R_{r\theta r \theta}]  &=&\frac{1}{f}[s^2,c^2,1]\,, \nonumber\\
{}[R_{\phi\psi\phi\psi},R_{\theta\psi \theta \psi},R_{\theta\phi \theta\phi}]  &=& M^2[s^2 c^2,c^2,s^2] \,.
\eea
Relations among these components are also evident. For example, the first line implies the tracefree property of the electric part of the Riemann tensor, $R_{\alpha\mu\beta\nu}(\partial_t)^\alpha (\partial_t)^\beta=0$ with respect to the observers at rest with respect to the coordinates or static observers, see below; the second line can be rewritten as
\beq
R_{r\phi r \phi}+R_{r\psi r \psi}=R_{r\theta r \theta}\,,
\eeq
and is a consequence of the first-kind Bianchi identities; the third line becomes
\beq
R_{\phi\psi\phi\psi}=R_{\theta\psi \theta \psi} R_{\theta\phi \theta\phi}\,,
\eeq
and is an expression of the Meusnier theorem concerning the relations among the curvatures of the $\theta$=const., $\phi$=const, and $\psi$=const. lines.

We can choose a family of fiducial observers as the static observes (at fixed spatial position $r=r_0,\theta=\theta_0,\phi=\phi_0, \psi=\psi_0$), i.e., with five velocity
\beq
u=\frac{1}{\sqrt{f}}\partial_t\,.
\eeq

These observers, in order to remain at a fixed spatial position, need to be radially accelerated to contrast the gravitational attraction from the hole, with acceleration
\beq
a(u)=\frac{M^2}{r^3}\partial_r\,,
\eeq 
where one should note the exponent 3 of the radial variable as a modification of the classical, 4d, Newton's law corresponding to the exponent 2.
A natural spatial frame associated with the family of static observers  is attached to the spherical coordinate lines
\bea
e_r&=& \frac{r}{\sqrt{r^2-M^2}}\partial_r\,, \nonumber\\
e_\theta &=& \frac{1}{r}\partial_\theta\,,\quad
e_\phi = \frac{1}{r\sin \theta }\partial_\phi\,,\quad
e_\psi = \frac{1}{r\cos \theta }\partial_\psi \,.
\eea

As in the $d=4$ case one defines the electric part (with respect to $u$) of the Riemann tensor as the following spatial (with respect to $u$), symmetric and tracefree rank-two tensor
\bea
E(u)_{\alpha\beta}&=& R_{\alpha \mu \beta \nu}u^\mu u^\nu\nonumber\\
&=& \frac{M^2}{r^2}{\rm diag}\left[0,-\frac{3}{r^2f},1,s^2, c^2\right]\,.
\eea
The magnetic part, is usually represented by the spatial (with respect to $u$) 3-tensor
\beq
H(u)_{abc}=\frac12 \eta_{abdef} R^{ef}{}_{cg}u^d u^g\,,
\eeq
which in this case vanishes identically (here we have denoted the indices of $H(u)$ with latin letters to remind their spatial character with respect to $u$). Hence, one is left with the so called \lq\lq mixed part" (purely spatial part) of the Riemann tensor deducible from Eqs. \eqref{compo}.

The existence of the Killing vectors $\partial_t$, $\partial_\phi$ and $\partial_\psi$ imply the occurrence of three constants of the motion along geodesics: $E$, $L_\phi$ and $L_\psi$ (denoted by an hat when measured in units of the mass $\mu$ of the probe, e.g., $\hat E=E/\mu$, etc.), respectively, so that
\bea
U^t&=&\dot t=\frac{\hat E}{\left(1-\frac{M^2}{r^2}\right)}\,, \nonumber\\
U^\phi&=&\dot \phi = \frac{\hat L_\phi}{r^2 \sin^2\theta }\,,\nonumber\\
U^\psi&=&\dot \psi = \frac{\hat L_\psi}{r^2 \cos^2\theta }\,,
\eea
where $U^\alpha(\tau)=\frac{dx^\alpha}{d\tau}\equiv \dot x^\alpha$ denotes the corresponding timelike five velocity parametrized by the proper time $\tau$, $U\cdot U=-1$. Here, a dot denotes differentiation with respect to $\tau$.

The $r$ and $\theta$ equations can be conveniently obtained within the Hamiltonian formalism (particularly useful when attempting a separation of variables). Let us define
\beq
H= g^{\mu\nu}P_\mu P_\nu=-\mu^2\,,
\eeq
where $P^\alpha=\mu U^\alpha$ is the four momentum of the mass $\mu$ particle (i.e., the \lq\lq probe").
Hamilton's equations can be separated through the introduction of a Carter-like constant $K$ as follows \cite{Chandrasekhar:1985kt}

\bea\label{QRQA}
P_r&=&  
\pm\frac{\sqrt{K^2 (M^2 - r^2) + M^2 r^2 \mu^2 + 
 r^4 (E^2 - \mu^2)} }{r^2-M^2} \,,\nonumber\\
P_\theta &=&  
\pm \frac{\sqrt{K^2-L_\phi^2 \csc^2\theta -L_\psi^2 \sec^2\theta}}{r^2}\,,
\eea

In the case $L_\psi=0$  (implying $\psi(\tau)=\psi_0$)  equatorial motion $\theta=\frac{\pi}{2}$ is allowed, with $K^2=L_\phi^2$, 
\bea
\dot t&=& \frac{\hat E}{\left(1-\frac{M^2}{r^2}\right)}>0 \nonumber\\
\dot \phi &=& \frac{\hat L_\phi}{r^2} \,,
\eea
and
\beq
 \dot r =\pm\frac{1}{\mu r^2} \sqrt{L_\phi^2 (M^2 - r^2) + M^2 r^2 \mu^2 + 
 r^4 (E^2 - \mu^2)}\,. 
\eeq

\subsection{Circular constant $\theta$ (non-equatorial) timelike orbits}

Timelike orbits at $r=r_0$ and $\theta=\theta_0$ exist for
\bea
\hat E^2 &=& \frac{(r_0^2-M)^2}{ r_0^2 (r_0^2-2 M^2)},\nonumber\\ 
\hat L_\phi^2 &=&  \frac{r_0^2 M^2}{ (r_0^2-2M^2)}\,\sin^4(\theta_0), \nonumber\\ 
\hat L_\psi^2 &=& \frac{r_0^2 M^2}{ (r_0^2-2M^2)}\,\cos^4 (\theta_0)\,,
\eea
namely for $r_0>\sqrt{2}M$ and any $\theta_0\in (0,\pi/2)$ with the two limiting cases: $\theta_0=0$
\bea
\hat L_\phi^2 &=&0\,,\quad
\hat L_\psi^2 =\frac{r_0^2 M^2}{ (r_0^2-2M^2)}\,,
\eea
and $\theta_0=\pi/2$ (corresponding to exchange the values of $\hat L_\phi$ and $\hat L_\psi$ of the case $\theta_0=0$), i.e.,
\bea
\hat L_\phi^2 &=& \frac{r_0^2 M^2}{ (r_0^2-2M^2)}, \quad
\hat L_\psi^2 = 0\,,
\eea
treated separately.

\subsection{Circular equatorial timelike orbits}

In addition to  $\hat L_\psi=0$ (i.e., constant $\psi$ along the motion, $\psi(\tau)=\psi_0$) and the equatorial motion limiting condition $\theta=\frac{\pi}{2}$ mentioned above, one can look for the existence of circular orbits, i.e., orbits at  $r=r_0$ such that
\beq
\phi(\tau)=\Omega_\tau \tau +\phi_0\,,\qquad t(\tau)=\Gamma \tau +t_0\,,
\eeq
with
\beq
\Omega_\tau=\frac{\hat L_\phi}{r_0^2}\,,\qquad \Gamma=\frac{\hat E}{\left(1-\frac{M^2}{r_0^2}\right)} \,,
\eeq
and
\bea
\label{circ_radii}
\hat E^2-\left(1+ \frac{\hat L_\phi^2}{r_0^2}\right)\left(1-\frac{M^2}{r_0^2} \right)=0 \,.
\eea
Eq. \eqref{circ_radii}  can be solved for $E$ and $L_\phi$ as functions of $r_0$
\bea
\hat E=\frac{E}{\mu}&=&      \frac{r_0^2-M^2}{r_0\sqrt{r_0^2-2M^2}}\,, \nonumber\\
\frac{\hat L_\phi}{M}=\frac{L_\phi}{M\mu} &=&  \frac{r_0}{\sqrt{r_0^2-2M^2}}\,.
\eea

\subsection{Unbound (hyperboliclike) orbits: the scattering angle}

In the case of equatorial motion, one can focus on unbound orbits, conveniently characterized by the scattering angle. To this aim  let us consider the equation
\beq
\label{dphi_dr}
\frac{d\phi}{dr}=\pm \frac{ L_\phi }{\sqrt{L_\phi^2 (M^2 - r^2) + M^2 r^2 \mu^2 + 
 r^4 (E^2 - \mu^2)}}\,.
\eeq
Eq, \eqref{dphi_dr} can be integrated exactly in terms of elliptic functions. In fact,  upon replacing $L_\phi$ by $b=\frac{L_\phi}{\sqrt{E^2-1}}$ (with $\mu=1$ without any loss of generality), one finds
\bea
\Delta \phi&=&\int_{r_1}^\infty dr \frac{b}{\sqrt{(r^2-r_1^2)(r^2-r_2^2)}}\nonumber\\
&=& \frac{b}{r_1}{\rm EllipticK}\left(\frac{r_2^2}{r_1^2}\right)\nonumber\\
&=& \frac{b}{r_1} \frac{\pi}{2}{}_2F_1\left[\frac12 , \frac12, 1, \frac{r_2^2}{r_1^2}\right]\,,
\eea
with $\hat b=b/M$, $p_\infty^2=E^2-1$  and
\bea
&&r_1=r_+,\quad r_2=r_-,\quad r_3=-r_2,\quad r_4=-r_1\,,\nn\\
&&r_\pm=\frac{M}{\sqrt{2}p_\infty}\sqrt{\hat{b}^2p_\infty^2{-}1\pm \sqrt{1{+}\hat{b}^2p_\infty^2(p_\infty^2(\hat{b}^2{-}4){-}2)}}\,.\nn\\
\eea
The turning points in the large impact parameter expanded form read
\bea\label{turns}
\frac{r_1}{M}&=& \hat b -\frac{1}{\hat b}\left(\frac12 +\frac{1}{2p_\infty^2}\right) +\frac{1}{\hat b^3}\left(-\frac58   - \frac{1}{8 p_\infty^4} - \frac{3}{4 p_\infty^2}\right)\nonumber\\
&+& O\left(\frac{1}{\hat b^4}\right)\,,\nonumber\\
\frac{r_2}{M}&=& 1 + \left(\frac12  + \frac{1}{2 p_\infty^2}\right)\frac{1}{\hat b^2} 
+O\left(\frac{1}{\hat b^4}\right)\,.\nn\\
\eea

It is straightforward to show that the Schwarzschild-Tangherlini $d=5$ scattering angle, $\chi=2\Delta \phi-\pi$, 
can be resummed as follows
\bea
\chi^{\rm ST 5}
&=& \pi\Big[-1+\sum_{n=0}^\infty \frac{f_n(b)}{p_\infty^{2n}}\Big]\,,
\eea
with
\beq
f_n(b)=\frac{(2n)!}{(n!)^2(2b)^{2n} }{}_2F_1\left(\frac{1}{4}+\frac{n}{2},\frac{3}{4}+\frac{n}{2},1,\frac{4}{b^2}\right)\,,
\eeq
namely, as a power series in $b$, i.e.,  in PM sense 
\bea
\chi^{\rm ST 5}
&=&\pi \left[\frac{\tilde \chi_2(p_\infty)}{b^2}+\frac{\tilde \chi_4(p_\infty)}{b^4}+\frac{\tilde \chi_6(p_\infty)}{b^6}\right.\nonumber\\
&+& \left. O\left(\frac{1}{b^8}\right)\right]\,,
\eea
that is, explicitly, 
\begin{widetext}
\bea\label{ang5d}
\chi^{\rm ST 5}
&=&\pi \left[
\frac{1}{b^2} \left(\frac{1}{2 p_{\infty
   }^2}+\frac{3}{4}\right)+
\frac{1}{b^4} \left(\frac{15}{8 p_{\infty
   }^2}+\frac{3}{8 p_{\infty
   }^4}+\frac{105}{64}\right)+
\frac{1}{b^6} \left(\frac{945}{128 p_{\infty
   }^2}+\frac{105}{32 p_{\infty
   }^4}+\frac{5}{16 p_{\infty
   }^6}+\frac{1155}{256}\right) \right. \nonumber\\
&+& \frac{1}{b^8} \left(\frac{15015}{512 p_{\infty
   }^2}+\frac{10395}{512
   p_{\infty }^4}+\frac{315}{64
   p_{\infty }^6}+\frac{35}{128
   p_{\infty
   }^8}+\frac{225225}{16384}\right)\nonumber\\
&+& \frac{1}{b^{10}} \left(\frac{3828825}{32768 p_{\infty
   }^2}+\frac{225225}{2048
   p_{\infty
   }^4}+\frac{45045}{1024
   p_{\infty
   }^6}+\frac{3465}{512
   p_{\infty }^8}+\frac{63}{256
   p_{\infty
   }^{10}}+\frac{2909907}{65536}\right)\nonumber\\
&+&  \frac{1}{b^{12}} \left(\frac{61108047}{131072 p_{\infty
   }^2}+\frac{72747675}{131072
   p_{\infty
   }^4}+\frac{1276275}{4096
   p_{\infty
   }^6}+\frac{675675}{8192
   p_{\infty
   }^8}+\frac{9009}{1024
   p_{\infty
   }^{10}}+\frac{231}{1024
   p_{\infty
   }^{12}}+\frac{156165009}{1048
   576}
\right)\nonumber\\
&+& \frac{1}{b^{14}} \left(\frac{3904125225}{2097152
   p_{\infty
   }^2}+\frac{1405485081}{524288
   p_{\infty
   }^4}+\frac{509233725}{262144
   p_{\infty
   }^6}+\frac{24249225}{32768
   p_{\infty
   }^8}+\frac{2297295}{16384
   p_{\infty
   }^{10}}+\frac{45045}{4096
   p_{\infty
   }^{12}}+\frac{429}{2048
   p_{\infty
   }^{14}}+\frac{2151252675}{419
   4304}\right)\nonumber\\
&+& \frac{1}{b^{16}} \left(\frac{62386327575}{8388608
   p_{\infty
   }^2}+\frac{105411381075}{8388
   608 p_{\infty
   }^4}+\frac{11712375675}{10485
   76 p_{\infty
   }^6}+\frac{11712375675}{20971
   52 p_{\infty
   }^8}+\frac{101846745}{65536
   p_{\infty
   }^{10}}+\frac{14549535}{65536
   p_{\infty
   }^{12}}+\frac{109395}{8192
   p_{\infty
   }^{14}}\right.\nonumber\\
&+& \left. \frac{6435}{32768
   p_{\infty
   }^{16}}+\frac{1933976154825}{
   1073741824}
\right)+\left. O\left(\frac{1}{b^{18}}\right)
\right]\,.
\eea
\end{widetext}
This expression can also be read as a power series in $p_\infty$ (with  coefficients depending on $b$), i.e.,  in PN sense, as follows
\bea
\chi^{\rm ST 5}
&=&\pi \left[\chi_0(b)+\frac{\chi_2(b)}{p_\infty^2}+ \frac{\chi_4(b)}{p_\infty^4}+\frac{\chi_6(b)}{p_\infty^6}\right.\nonumber\\
&+& \left. O\left(\frac{1}{p_\infty^8} \right) \right]\,,
\eea
where
\bea
\chi_0(b)&=&\frac{3}{4 b^2}+\frac{105}{64
   b^4}+\frac{1155}{256 b^6}+\frac{225225}{16384 b^8}+O\left(\frac{1}{b^{10}} \right)\,,\nonumber\\
\chi_2(b)&=& \frac{1}{2 b^2}+\frac{15}{8
   b^4}+\frac{945}{128 b^6}+\frac{15015}{512 b^8}+O\left(\frac{1}{b^{10}} \right)\,,
\nonumber\\
\chi_4(b)&=&\frac{3 }{8 b^4}+\frac{105 }{32 b^6}+\frac{10395 }{512 b^8} +O\left(\frac{1}{b^{10}} \right)\,,\nonumber\\
\chi_6(b)&=& \frac{5 }{16 b^6} +\frac{315 }{64 b^8} +O\left(\frac{1}{b^{10}} \right)\,,
\eea
etc.

\section{Radial action}

Because of the well known relation  between the radial action and the scattering angle, 
in this subsection we will  (analytically) compute the  (regularized) radial action $I_r$ along scattering timelike geodesics, following Refs. \cite{Parnachev:2020zbr,Ivanov:2025ozg,Bini:2025ltr,Bini:2025bll}. The latter is defined as
\beq
I_r =\int_{\infty}^{r_1} P_r dr\,, 
\eeq
where the conjugate momentum to the radial coordinate is defined in Eq. \eqref{QRQA} and where $r_1$ is its the largest turning point as given  in Eq. \eqref{turns} above. The scattering angle is defined through the relation
\bea
\frac{\chi+\pi}{2}=-\frac{\partial I_r}{\partial L_\phi}\,.
\eea
Let's introduce the new variables
\beq
r=\frac{M\sqrt{2\bar{E}}}{jx},\quad L_\phi=\mu M j,\quad E=\mu \sqrt{1+2\bar{E}}\,,
\eeq
which allow us to rewrite the turning points as
\bea
x_1&=&1+\frac{\frac{1}{2}+\bar{E}}{j^2}+O\left(\frac{1}{j^4}\right)\,,\nn\\
x_2&=&\frac{j}{\sqrt{2\bar{E}}}-\frac{1+2\bar{E}}{2\sqrt{2\bar{E}}j}-\frac{1+12\bar{E}+20\bar{E}^2}{8\sqrt{2\bar{E}}j^3}+O\left(\frac{1}{j^4}\right)\,,\nn\\
x_3&=&-x_2\,,\nn\\
x_4&=&-x_1\,.
\eea
In order to select only the finite part of the radial action, it is enough to truncate $x_1$ at the first order in large j (i.e., Post-Minkowskian) expansion~\cite{Damour:1988mr}.

Consequently, the radial action integrand becomes
\bea\label{IrU}
I_r=M \mu\, {\mathcal I}_r\,,
\eea
with
\beq
\label{IrUbis}
{\mathcal I}_r=\int_0^1 dx \frac{j  \sqrt{1-x^2}\sqrt{1-\frac{x^2(1+2\bar{E}x^2)}{j^2(x^2-1)}}}{x^2\left(1-\frac{2\bar{E}x^2}{j^2}\right)}\,,
\eeq
and then
\bea
\frac{\chi+\pi}{2}=-\frac{\partial {\mathcal I}_r}{\partial j}\,.
\eea

Introducing the notation 
\beq
\bar{E}=\frac{p_\infty^2}{2},\quad j=\frac{2p_\infty}{y}\,,
\eeq
the integral \eqref{IrUbis} can be computed expanding in   
small $y$, integrating over $x$, and then resumming the expression obtained in $y$.
The final result (involving some technicalities as explained for example in Refs. \cite{Bini:2025ltr,Bini:2025bll}) is 
\bea
{\mathcal I}_r&=& -  \pi^2 \sum_{k} \frac{p_\infty^{1 - 2 k}}{\Gamma (2 k+1)}(-16)^k\left[
\frac{\Gamma\left(\frac34\right)}{y} \,F_1\right.\nonumber\\
&+&\left. \frac{[3+4k(k-2)]}{64}\,y \Gamma\left(\frac54\right) F_2 \right]\,,
\eea
with $F_1$ and $F_2$ the following regularized hypergeometric functions 
\begin{widetext}
\bea
F_1&=& _6\tilde{F}_5\left(\frac{1}{4},1,\frac{5}{8}-\frac{k}{4},\frac{7}{8}-\frac{k}{4},\frac{9}{8}-\frac{k}{4},\frac{11}{8}-\frac{k}{4};\frac{5}{4},1-\frac{k}
   {2},1-\frac{k}{2},\frac{3}{2}-\frac{k}{2},\frac{3}{2}-\frac{k}{2};y^4\right)\,,\nonumber\\
F_2&=& _6\tilde{F}_5\left(-\frac{1}{4},1,\frac{1}{8}-\frac{k}{4},\frac{3}{8}-\frac{
   k}{4},\frac{5}{8}-\frac{k}{4},\frac{7}{8}-\frac{k}{4};\frac{3}{4},\frac{1}{2
   }-\frac{k}{2},\frac{1}{2}-\frac{k}{2},1-\frac{k}{2},1-\frac{k}{2};y^4\right)\,.
\eea
\end{widetext}

The first few terms (from $k=0$ to $k=10$) of this sum once re-expanded in series of $y$ up to $O(y^{10})$ read
\bea
\frac{\chi}{\pi}&=&
\frac{\frac{3
   p_\infty^2}{4}+\frac{1}{2}}{j^2}\nonumber\\
&+&\frac{\frac{105
   p_\infty^4}{64}+\frac{15
   p_\infty^2}{8}+\frac{3}{8}}{j^4}\nonumber\\
&+&\frac{\frac{1155
   p_\infty^6}{256}+\frac{945 p_\infty^4}{128}+\frac{105
   p_\infty^2}{32}+\frac{5}{16}}{j^6}\nonumber\\
&+&\frac{\frac{225225
   p_\infty^8}{16384}+\frac{15015 p_\infty^6}{512}+\frac{10395
   p_\infty^4}{512}+\frac{315
   p_\infty^2}{64}+\frac{35}{128}}{j^8}\nonumber\\
&+& O(\frac{1}{j^{10}})\,,
\eea
and can be used to check the expression for the scattering angle shown above.

\subsection{Geodesic deviations}

In view of characterizing differences with respect the more familiar $4d$ Schwarzschild spacetime, 
in this subsection we will discuss the deviations from (timelike, equatorial) circular geodesics for timelike particles. Since this is a genuine five-dimensional geometry whose angular part is the geometry of the $S^3$ sphere, the geodesic deviation is expected to exhibit  some substantial differences with respect to the 4d Schwarzschild case. Actually, 
the STd5 case is quite different also from other 5d cases studied in the literature, as, for example,
the Topological Star spacetime \cite{Bini:2025qyn}. In the latter case, in fact, the angular part is an $S^2$ sphere, while the extra fifth dimension is compact. 

In order to characterize the geodesic deviation in the STd5 spacetime, let us consider as a reference geodesic ${\mathcal C}_c$ the circular geodesic studied above, with five-velocity
\beq
\label{geo_circ}
U_{\rm c}=\frac{r_0}{\sqrt{r_0^2-2M^2}}\left(\partial_t+\frac{M}{r_0^2}\partial_\phi\right) \,,
\eeq 
($\Omega_{\rm orb}=\frac{M}{r_0^2}$ being the orbital frequency) and parametric equations $x^\alpha =x_c^\alpha(\tau)$, i.e., 
\bea
t&=&\frac{r_0}{\sqrt{r_0^2-2M^2}} \tau\,,\quad r=r_0\,,\quad \theta=\frac{\pi}{2}\,,\nonumber\\
\phi&=&\frac{r_0}{\sqrt{r_0^2-2M^2}} \frac{M}{r_0^2} \tau\,,\quad \psi=0\,.
\eea
We will proceed by studying deviations from this curve, associated with a vector  $\eta^\alpha$ orthogonal to $U_{\rm c}$ ($\eta\cdot U_{\rm c}=0$), and undergoing the geodesic deviation transport law
\beq
\label{geo_dev}
\nabla_{U_{\rm c}}\nabla_{U_{\rm c}} \eta^\alpha +R^\alpha{}_{\beta\mu\nu}|_{{\mathcal C}_c} U_{\rm c}^\beta U_{\rm c}^\nu \eta^\mu=0\,.
\eeq
Note that $U_c$ is timelike  for $r_0> \sqrt{2}M$ (i.e., at $r_0=\sqrt{2}M$, the photonsphere of massless particles, $U_c$ becomes null). Therefore, the region of validity of the present analysis is $r_0> \sqrt{2}M$.

Let us parametrize by $s$ the (spatial) curve whose (not unitary) tangent vector is $\eta^\alpha$.
Eq. \eqref{geo_dev} is solved by
\bea
\eta^t(s)&=&\frac{B_\phi M r_0^2}{M^2-r_0^2}  (1-\cosh (\Omega s))\,, \nonumber\\
\eta^r(s)&=& -B_\phi M \sinh (\Omega s)\,, \nonumber\\
\eta^\theta(s)&=& \eta^\theta(0)   \cos \left(\frac{r_0}{2M}\Omega s \right)-\eta^\psi(0) \sin \left(\frac{r_0}{2M}\Omega s \right)\,, \nonumber\\
\eta^\phi(s)&=&-B_\phi (1-\cosh (\Omega s))\,, \nonumber\\
\eta^\psi(s)&=&  \eta^\psi(0) \cos \left(\frac{r_0}{2M}\Omega s \right) +\eta^\theta(0) \sin \left(\frac{r_0}{2M}\Omega s \right)\,,\nonumber\\
\eea
where $B_\phi$ is an integration constant such that $\frac{d\eta^r(s)}{ds}|_{s=0}=-B_\phi M\Omega $  and
\beq
\Omega=\frac{2M^2}{r_0^2\sqrt{r_0^2-2M^2}}
\eeq
is the analogous of the epicyclic frequency (and different from the orbital frequency $\Omega_{\rm orb}$).
In general, the evolution of $\eta^\alpha$ corresponds to both a spatial  rotation with frequency $\frac{r_0}{2M}\Omega$ in the (elliptic) $\theta$ and $\psi$ subspace, and to a boost in the 
the (hyperbolic) $r$ and $\phi$ subspace with frequency $\Omega$, the $t$ component following from the orthogonality of $U_c^\alpha$ and $\eta^\alpha$. In fact, referring for simplicity to coordinate components, we find
\beq
\begin{pmatrix}
\eta^\theta(s)\\
\eta^\psi(s)
\end{pmatrix}
= 
\begin{pmatrix}
\cos(\frac{r_0}{2M}\Omega s) &-\sin(\frac{r_0}{2M}\Omega s)  \\
\sin(\frac{r_0}{2M}\Omega s) &\cos(\frac{r_0}{2M}\Omega s)\\
\end{pmatrix}
\begin{pmatrix}
\eta^\theta(0)\\
\eta^\psi(0) 
\end{pmatrix}\,,
\eeq
and
\bea
&&\begin{pmatrix}
-\frac{\eta^r(s)}{M}\\
\eta^\phi(s)+B_\phi\\
\end{pmatrix}
=
\begin{pmatrix}
\cosh (\Omega s) &\sinh (\Omega s)\\
\sinh (\Omega s) & \cosh (\Omega s)\\
\end{pmatrix}
\begin{pmatrix}
0\\
B_\phi \\
\end{pmatrix}\,.
\nonumber\\
\eea
At the values $s=s_k$, with
\beq
s_k=2k\pi \frac{2M}{\Omega r_0}\,,\quad k\in {\mathbb Z}\,,
\eeq
the rotation  in the $\theta$ and $\psi$ subspace reduces to the identity, and the evolution gives back the initial values.
We recall that this is a quite different behavior with respect to the standard $4d$ Schwarzschild spacetime, where only rotations are involved. Indeed, in the 4d Schwarzschild spacetime the circular orbit four velocity reads
\beq
U_{\rm cs}=\frac{1}{\sqrt{1-\frac{3M}{r_0}}}\left(\partial_t+\sqrt{\frac{M}{r_0^3}}\partial_\phi\right) \,,
\eeq 
so that $U_{\rm cs}$ is timelike for $r_0>3M$, i.e.,  outside the light ring located at $r_0=3M$.
The component of the deviation vector along the $\theta$ direction  oscillates with frequency $\frac{\sqrt{M}}{r_0 \sqrt{r_0-3M}}$ as
\beq
\eta^\theta(s)= A_\theta \cos \left(\frac{\sqrt{M} s}{r_0 \sqrt{r_0-3M}} \right)+B_\theta \sin \left(\frac{\sqrt{M} s}{r_0 \sqrt{r_0-3M}} \right)\,,
\eeq
while all the other components oscillate with the epicyclic frequency
\beq
\Omega_{\rm ep}=\sqrt{\frac{M}{r_0^3}}\sqrt{\frac{r_0-6M}{r_0-3M}}\,,
\eeq
as
\bea
\eta^r(s)&=&  -\frac12 B_\phi \sqrt{r_0(r_0-6M)}\cos (\Omega_{\rm ep}s)+B_r\sin (\Omega_{\rm ep}s)\,,\nonumber\\
\eta^\phi(s)&=& C_\phi   +\frac{2 B_r}{ \sqrt{r_0(r_0-6M)}}\cos (\Omega_{\rm ep}s)+B_\phi\sin (\Omega_{\rm ep}s) \,,\nonumber\\
\eea
with $B_r$ and $B_\phi$ related to $\eta^r(0)$ and $\eta^\phi(0)$ as
\bea
\eta^r(0)&=&  -\frac12 B_\phi \sqrt{r_0(r_0-6M)}\,, \nonumber\\
\eta^\phi(0)&=& C_\phi   +\frac{2 B_r}{ \sqrt{r_0(r_0-6M)}} \,, 
\eea
while $\eta^t(s)$ follows from the orthogonality condition between $U_{\rm cs}^\alpha$ and $\eta^\alpha$ simply related to $\eta^\phi(s)$,
\beq
\eta^t(s)=\frac{\sqrt{M}r_0^{3/2}}{r_0-2M}\eta^\phi(s)\,.
\eeq
Noticeably, also in this case the evolution can be written in terms of a rotation  matrix.

Besides geodesic deviation, another geometrically relevant characterization of ST5d concerns transport laws of vectors along special curves, see, e.g. \cite{Bini:2004sy}.
To this end let us study the parallel transport law of a vector $X$ along the circular orbit of Eq. \eqref{geo_circ},
\beq
\nabla_{U_{\rm c}}X=0\,.
\eeq
We find that the $\theta$ and $\psi$ components are constants 
\beq
X^\theta=X^\theta(0)\,,\quad X^\psi=X^\psi(0)\,,
\eeq
while the remaining components are evolving according the following (coupled) system
\bea
\label{sys_par}
\frac{dX^t}{d\tau } &=& -\frac{M^2}{ (r_0^2-M^2)(r_0^2-2M^2)^{1/2}}X^r\,, \nonumber\\
\frac{dX^r}{d\tau } &=& \frac{M (r_0^2-M^2)}{r_0^2(r_0^2-2M^2)^{1/2}} X^\phi -\frac{M^2 (r_0^2-M^2) }{ r_0^4(r_0^2-2M^2)^{1/2}}X^t\,,\nonumber\\
\frac{dX^\phi}{d\tau } &=& -\frac{M}{ r_0^2(r_0^2-2M^2)^{1/2}} X^r \,.
\eea
The solution of Eqs. \eqref{sys_par} is the following
\bea
X^t  &=&  \frac{C_1}{\Omega_{\rm orb}} -\frac{r_0 \Omega_{\rm orb} \zeta  (\tau)}{(\Omega_{\rm orb}^2r_0^2-1) ( 1-2\Omega_{\rm orb}^2r_0^2)^{1/2}}  \,, \nonumber\\
X^r  &=&  \Sigma (\tau)  \,,\nonumber\\
X^\phi &=&  C_1+\frac{\zeta(\tau) }{r ( 1-2\Omega_{\rm orb}^2r^2)^{1/2} } \,,
\eea
having defined
\bea
\zeta(\tau)&=& C_2\cos(\Omega_{\rm orb} \tau )-  C_3\sin(\Omega_{\rm orb} \tau)\,,\nonumber\\
\Sigma (\tau)&=& C_2\sin(\Omega_{\rm orb} \tau) +C_3\cos(\Omega_{\rm orb} \tau)\,,
\eea
and recalling that $\Omega_{\rm orb} =\frac{M}{r_0^2}$ and $C_2=\zeta(0)$, $C_3=\Sigma (0)$.
A special choice of the  integration constants can be used to identify a spatial frame parallely propagated then along $U_{\rm c}$.

Here, contrary to initial expectations, we see that the fifth dimension does not play a special role and the situation is quite similar to that of the standard 4d Schwarzschild case. Even if this behaviour can be mainly understood as a consequence of the choice of $\theta=\pi/2$ for the parametric equations of the reference curve,  
the question arises if some other geometrically motivated transport law may allow for any role to  be played by the fifth dimension.
It is then natural to study the Fermi-Walker transport along an accelerated (for example, a circular orbit $U$ in the family of $U_{\rm c}$). Vectors undergoing Fermi-Walker transport along a timelike curve with unit tangent vector $U$ ($U\cdot U=-1$) correspond  to test gyroscopes carried by the observers with five velocity $U$.
To this end let us consider a generic (non geodesic) circular equatorial orbit 
\beq
U=\Gamma (\partial_t +\Omega \partial_\phi)\,,
\eeq
with $\Omega$ unspecified and
\beq
\Gamma=\frac{r_0}{\sqrt{r_0^2-M^2-\Omega^2r_0^4}} \,.
\eeq
This orbit is accelerated with acceleration directed radially
\beq
a(U)=\nabla_UU =\frac{(r_0^2-M^2) (M^2-\Omega^2 r_0^4)}{ (r_0^2- M^2-\Omega^2 r_0^4) r_0^3} \partial_r\,. 
\eeq
The Fermi-Walker transport equations for a generic vector $X$
\beq
\nabla_U X^\mu-[U\wedge a(U)]^\mu{}_\nu X^\nu=0
\eeq
imply oscillations of all its components with the Fermi-Walker frequency
\beq
\Omega_{\rm fw}=\Omega\, \frac{(r_0^2-2M^2)}{( r_0^2-M^2 \Omega^2 r_0^4)}\,. 
\eeq
The explicit solution reads
\bea
X^t &=& -\frac{\Omega r_0^4}{ (r_0^2-M^2) (r_0^2- M^2-\Omega^2 r_0^4)^{1/2}} \Sigma_{\rm fw}(\tau) + C_1\,,\nonumber\\
X^r &=& \zeta_{\rm fw}(\tau)\,, \nonumber\\
X^\phi &=& -\frac{\Sigma_{\rm fw}(\tau)}{(r_0^2- M^2-\Omega^2 r_0^4)^{1/2}}+\Omega C_1\,,
\eea
having defined
\bea
\zeta_{\rm fw}(\tau)&=& C_2\cos(\Omega_{\rm fw} \tau )-  C_3\sin(\Omega_{\rm fw} \tau)\,,\nonumber\\
\Sigma_{\rm fw}(\tau)&=& C_2 \sin(\Omega_{\rm fw}\tau )+C_3\cos(\Omega_{\rm fw}\tau )\,,
\eea
while
\beq
X^\theta=X^\theta(0)\,,\quad X^\psi=X^\psi(0)\,,
\eeq
as before.
We have then shown that while the Fermi-Walker transport law  simply corresponds to a rotation from initial data, it does not allow for any special role of the extra dimension associated with the coordinate $\psi$.

\subsection{QNM: analytic expressions in eikonal limit}

In the context of geodesic motion, another quantity which is worth to be extracted and studied in the STd5 case is the so-called Lyapunov exponent, which characterizes the instability of null geodesics in the vicinity of the (unstable) photonsphere \cite{Cardoso:2008bp,Bianchi:2021mft}. 
The latter will be used to compute the quasinormal modes (QNM) spectrum.

Let us start with the equatorial radial effective potential for timelike geodesics, given in Eq. \eqref{QRQA}. Equivalently,  
\bea\label{geopr}
P_r^2\equiv Q_{\rm r,geo}{=}\frac{r^4 (E^2 {-} \mu^2){+}L_\phi^2 (M^2{-}r^2){+} M^2 r^2 \mu^2  }{(r^2-M^2)^2}\,.\nn\\
\eea
The critical circular geodesic is defined as
\beq
Q_{\rm r,geo}(r,L_\phi)=0=\partial_r Q_{\rm r,geo}(r,L_\phi)\,.
\eeq
These two equations can be solved for $r$ and $E$, i.e., the radius and the energy of the critical orbit, respectively. Their explicit expressions are
\bea
r_c&=&\frac{\sqrt{2}L_\phi M}{\sqrt{L^2_\phi-M^2\mu^2}}\,,\nn\\
E_c&=&\frac{L_\phi}{2M}+\frac{M\mu^2}{2L_\phi}\,.
\eea
Recalling that a dot denotes differentiation with respect the proper time, with
\bea
\frac{E}{\mu}=\left(1-\frac{M^2}{r^2}\right)\dot t\quad,\quad \frac{P_r}{\mu}=\frac{\dot r}{1-\frac{M^2}{r^2}}\,,
\eea
and using Eq. \eqref{geopr}, one writes
\bea
\frac{dr}{dt}&=&\frac{(r^2-M^2)}{r^4E}\nn\\
&\times&\sqrt{(E^2-\mu^2)r^4+L_\phi^2(M^2-r^2)+M^2\mu^2r^2}\,.\nn\\
\eea
Approaching the photonsphere, $r\to r_c$, one finds  the following asymptotic behavior
\beq
\frac{dr}{dt}\sim
-\lambda (r-r_c)\,,
\eeq
where the Lyapunov exponent $\lambda$ turns out to be given by 
\beq
\lambda=\frac{(L_\phi^2-M^2\mu^2)^\frac{3}{2}}{\sqrt{2}L_\phi^3M}\,.
\eeq
Identifying $L_\phi$ with the angular quantum number $\ell$ (in units of $\hbar =1$; properly speaking $L_\phi=\hbar \ell$), in the eikonal limit one obtains an analytic expression for the QNM frequencies: the real part is given by the orbital frequency of the probe on the unstable circular orbit, while the imaginary part is determined by the corresponding Lyapunov exponent
\bea
\omega_n^{\rm ST}&=&\frac{1}{2M}\left[ \ell +\frac{M^2\mu^2}{ \ell}-i\sqrt{2} \left(n+\frac12 \right)\right.\nonumber\\
&  +& \left. O\left( \frac{1}{\ell^2}\right)\right]\,,
\eea
where in the imaginary part appears and integer number $n$ usually called the overtone number. In $d=4$ the analogous expression (always in units of $\hbar =1$) is 
\bea
\omega_n^{\rm S}&=& \frac{1}{3\sqrt{3}M}\left[\ell+\frac12+\frac{-7+108 M^2\mu^2}{24\ell}\right.\nonumber\\
& -& \left. i \left(n+\frac12\right)  +O\left( \frac{1}{\ell^2}\right)\right]\,,
\eea
where  $M\mu \sim M\mu \frac{G}{c\hbar}$ is dimensionless.
Recalling that in the standard 4d case $b_{\rm crit}=3\sqrt{3}$ while now $b_{\rm crit}=2$, in both cases we find $b_{\rm crit} M\omega_n\sim \ell$ plus corrections $O(\ell^0)$. The results provided by the eikonal estimation are tested against a more robust numerical integration procedure as displayed in Table \ref{tabQNM}.

\begin{table}[]
\caption{\label{tabQNM}  QNM frequencies computed numerically and in the eikonal  approximation limit. It is worth to recall  that the eikonal limit reproduces better and better the QNM only for large $\ell$. The numerical integration is based on the requirement of vanishing the Wronskian computed starting from the solutions with ingoing boundary conditions at the horizon and outgoing at infinity as discussed as an example in \cite{Berti:2009wx,Pani:2013pma,Cardoso:2014sna}.} 
\begin{tabular}{|c|c|c|}
\hline
   $\ell$       & Eikonal            & Numerical               \\ \hline
$0$  & $$                 & $0.536252 - 0.376968 i$ \\ \hline
$1$  & $0.5 - 0.353553 i$ & $1.01602 - 0.362324 i$  \\ \hline
$2$  & $1. - 0.353553 i$  & $1.51057 - 0.357535 i$  \\ \hline
$3$  & $1.5 - 0.353553 i$ & $2.00789 - 0.355801 i$  \\ \hline
$4$  & $2. - 0.353553 i$  & $2.50629 - 0.354993 i$  \\ \hline
$5$  & $2.5 - 0.353553 i$ & $3.00523 - 0.354553 i$  \\ \hline
$6$  & $3. - 0.353553 i$  & $3.50448 - 0.354287 i$  \\ \hline
$7$  & $3.5 - 0.353553 i$ & $4.00392 - 0.354115 i$  \\ \hline
$8$  & $4. - 0.353553 i$  & $4.50348 - 0.353997 i$  \\ \hline
$9$  & $4.5 - 0.353553 i$ & $5.00313 - 0.353912 i$  \\ \hline
$10$ & $5. - 0.353553 i$  & $5.50285 - 0.35385 i$   \\ \hline
\end{tabular}
\end{table}

\section{Massless scalar wave equation}

The Klein-Gordon equation describing scalar massless particles 
\beq
\Box \psi=g^{\mu\nu}\nabla_\mu \partial_\nu \psi =0\,,
\eeq
can be separated through the ansatz
\bea
\Phi(t,r,\theta,\phi,\psi)&=&\sum_{\ell,m_\phi,m_\psi}e^{im_\phi \phi+i m_\psi \psi}S_{\ell m_\phi m_\psi}(\theta)\times \nonumber\\
&& \int \frac{d\omega}{2\pi}e^{-i\omega t}R_{\ell m_\phi m_\psi \omega}(r)\,,
\eea
where the angular equation corresponds to the equation for spherical harmonics in five dimensions (on $S_3$)
\bea
&&\Big[\frac{1}{\sin\theta \cos\theta }\partial_\theta \Big(\sin\theta \cos \theta  \partial_\theta\Big)-\frac{m_\phi^2}{\sin^2\theta}-\frac{m_\psi^2}{\cos^2\theta}\Big]S(\theta)\nn\\
&=&-\ell(\ell+2)S(\theta)\,,
\eea
where we have denoted $S_{\ell m_\phi m_\psi}(\theta)=S(\theta)$ for simplicity. 

The solution of this equation reads \cite{Berti:2005gp}
\bea
S(\theta)=\cos^{|m_\psi|}(\theta)\sin^{|m_\phi|}(\theta){}_2F_1 \left(\left[a,b\right],\left[c\right],\cos^2(\theta)\right)\,,\nn\\
\eea
where 
\bea
a&=&\frac{ m_\phi+m_\psi-\ell}{2}\,, \nonumber\\
b&=& \frac{m_\phi+m_\psi+\ell}{2} +1\,, \nonumber\\
c&=& 1+m_\psi\,.
\eea

The radial differential equation reads
\bea
\label{eqR}
&&(r^2-M^2)R''(r)+\left(3r-\frac{M^2}{r}\right)R'(r)\nn\\
&&\quad +\frac{r^4\omega^2-(r^2-M^2)\ell(\ell+2)}{r^2-M^2}R(r)=0\,.
\eea
Introducing the notation
\beq
{\mathcal L}=\ell(\ell+2)\,,
\eeq
the transformation to the normal (or canonical form) is obtained by the replacement
\be
R(r)=\frac{\psi(r)}{\sqrt{r(r^2-M^2)}}\,,
\ee
so that Eq. \eqref{eqR}  becomes
\bea
\label{HE_normal}
\psi''(r)+Q_W(r)\psi(r)=0\,,
\eea
where
\bea
Q_W(r)&=&\frac{4\omega^2r^6{-}3r^4{-}4{\mathcal L}r^2(r^2{-}M^2){+}6M^2r^2{+}M^4}{4r^2(r^2-M^2)^2}\,.\nn\\
\eea
Equivalently,
\bea
Q_W(r)&=&\omega^2 +\frac{1}{4r^2}+\frac{3+2{\mathcal L}-3M^2\omega^2}{4M(r+M)}\nonumber\\
&+&\frac{3M^2\omega^2 -3-2{\mathcal L}}{4M(r-M)}+\frac{1+M^2\omega^2}{4(r+M)^2}\nonumber\\
&+& \frac{1+M^2\omega^2}{4(r-M)^2}\,,
\eea
where the regular singular points are located at $r=0,\pm M$ while $r\to \infty$ corresponds to an irregular singular point.

The Frobenius exponents describing the leading behavior near the Fuchsian singularities are
\bea
&&\psi(r)\underset{r\to 0}{\sim}r^{\alpha_0},\quad \alpha_0=\frac{1}{2}\,,\nn\\
&&\psi(r)\underset{r\to M}{\sim}(r- M)^{\alpha^{+}_{M,\pm}},\quad \alpha^{+}_{M,\pm}=\frac{1}{2}(1\pm iM\omega)\,,\nn\\
&&\psi(r)\underset{r\to -M}{\sim}(r+ M)^{\alpha^{-}_{M,\pm}},\quad \alpha^{-}_{M,\pm}=\frac{1}{2}(1\pm iM\omega)\,,\nn\\
\eea
where the  exponent $\alpha^{+}_{M,-}$ corresponds to the requirement of ingoing boundary conditions at the horizon $r=M$.

The behavior near the irregular singularity is transcendental rather than power-like 
\beq
\psi(r)\underset{r\to\infty}{\sim}e^{\pm i\omega r}\,,
\eeq
where the $+$ sign correspond to outgoing waves at infinity.

Eq. \eqref{HE_normal} (already in its normal form) can be cast in the standard form of a Reduced Confluent Heun equation (RCHE) via the map
$r=M\sqrt{1+y}$ (see Refs. \cite{Bonelli:2022ten,Consoli:2022eey} and the discussion below).

Let us look for an \lq\lq in" type solution (purely ingoing at the horizon) and an \lq\lq up" solution (purely outgoing at infinity), both in PN sense.
Introducing the weights $M\to M\eta$ and $\omega \to \omega \eta$ (different from the $d=4$ case) and PN expanding the radial equation one finds the in solution in the form
\bea
R_{\rm in}(r)=r^{\nu-\ell}R_{\rm in, resc}(r)\,,
\eea
where $\nu=\nu(\ell)$ will be given below (see Eq. \eqref{a_l_generico}) one finds that $R_{\rm in, resc}(r)$ does not contain $\ln r$ anymore, and it is given by
\begin{widetext}
\bea
R_{\rm in, resc}(r)&=& r^\ell-\eta^2\frac{r^{\ell -2} \left(M^2 \ell
   ^2+2 M^2 \ell +r^4 \omega
   ^2\right)}{4 (\ell +2)}\nonumber\\
&+&\eta^4 \left[ 
\frac{r^{\ell -4}}{32 (\ell
   -1) (\ell +1)^2 (\ell +2) (\ell
   +3)} 
\left(M^4 \ell
   ^7+3 M^4 \ell ^6-7 M^4 \ell ^5-23
   M^4 \ell ^4-r^4 \omega ^2
   \left(72 M^2+r^4 \omega^2\right)\right.\right.\nonumber\\
&+&6\left.\left.
\ell ^3 \left(6 M^4+12
   M^2 r^4 \omega ^2+r^8 \omega
   ^4\right)
+\ell ^2 \left(44 M^4+48
   M^2 r^4 \omega ^2+r^8 \omega
   ^4\right)
+\ell  \left(24 M^4+12
   M^2 r^4 \omega ^2-r^8\omega^4
\right)
\right)
\right]\nonumber\\
&+& O(\eta^6)\,,
\eea
\end{widetext}
where we displayed the first terms of the PN expansion.

The PN up solution follows  by replacing $\ell\to -\ell -2$ in the in solution (as in the 4d Schwarzschild case was obtained by replacing $\ell\to -\ell -1$).

\section{Seiberg-Witten approach and check of the renormalized angular momentum $\nu$}

It is now well known that the radial equation \eqref{eqR} can be mapped to the quantum Seiberg-Witten (SW) curve with two non symmetric mass hypermultiplets $(0,2)$. For a detailed and updted reference of the topic, we refer to Refs. \cite{Aminov:2020yma,Bianchi:2021mft,Consoli:2022eey,Aminov:2023jve}. A very recent development of such kind of analytic technologies applied on black hole perturbation theory can be found in \cite{Fioravanti:2025bts}. The latter can be obtained from the $(1,2)$ curve, also known as the Confluent Heun equation (CHE), which, for example, describes gravitational perturbations in the Kerr-Newman geometry, Topological Star and W-soliton solutions \cite{Bianchi:2025ydq}. The $(1,2)$ curve can be written as
\bea
Q_{1,2}(y)&=&-\frac{Q^2}{4}+\frac{1-(m_1-m_2)^2}{4y^2}\nn\\
&+&\frac{1-(m_1+m_2)^2}{4(1+y)^2}-\frac{Q m_3}{y}\nn\\
&-&\frac{1-2(m_1^2+m_2^2)-2Q(1-m_1-m_2)+4u}{4y(1+y)}\,.\nn\\
\eea
In the SW perspective, the confluence corresponds to the decoupling of the hypermultiplet of mass $m_3$. Operationally, this is implemented by taking the double scaling limit $Q\to0$ and $m_3\to \infty$, while keeping the  product $Qm_3$ as fixed, and identifying it with the new gauge coupling $q=-Qm_3$. The resulting curve is
\bea\label{Q20}
Q_{0,2}(y)&=&\frac{1-(m_1-m_2)^2}{4y^2}+\frac{1-(m_1+m_2)^2}{4(1+y)^2}\nn\\
&+&\frac{1-2(m_1^2+m_2^2)+4u}{4(1+y)}\nn\\
&+&\frac{-1+2(m_1^2+m_2^2)+4(q-u)}{4y}\,,
\eea
through the introduction of the following dimensionless coordinate
\beq
\label{map_ry}
r=M\sqrt{1+y}\,.
\eeq
The dictionary for such a correspondence is the following
\be\label{dict}
m_1=-m_2=i\sqrt{q},\quad u=\left(\frac{\ell+1}{2}\right)^2-q\,,
\ee
with the gauge theory coupling
\beq
q=\frac{M^2\omega^2}{4}\,.
\eeq
Some brief reminders are in order.

In the non commutative Nekrasov-Shatashvili background~\cite{Nekrasov:2009rc},  the dynamics of the gauge theory is described by a quantum curve obtained from the classical one
\beq
q^2y^2P_L(x)+yP_0(x)+P_R(x)=0\,,
\eeq
after replacing the variables $y$ and $x$ with operators satisfying the commutation relation
\beq
[\hat x,\ln \hat y]=\hbar=1\,.
\eeq
A possible way to write the quantum curve is then given by
\beq\label{qSW}
\Big[qP_L\left(x-\frac{\hbar}{2}\right)\hat y+P_0(x)+P_R\left(x+\frac{\hbar}{2}\right)\hat y^{-1}\Big]\tilde{U}(x)=0\,.
\eeq
Introducing the functions
\bea
W(x)&=&\frac{1}{P_R\left(x+\frac{\hbar}{2}\right)}\frac{\tilde{U}(x)}{\tilde{U}(x+\hbar)}\,,\nn\\
M(x)&=&P_L\left(x-\frac{\hbar}{2}\right)P_R\left(x-\frac{\hbar}{2}\right)\,,
\eea
Eq. \eqref{qSW} can be cast in the form
\beq
qM(x)W(x)W(x-\hbar)+P_0(x)W(x)+1=0\,,
\eeq
which can be recursively solved considering small $q$
\bea
W(x)&=&-\frac{1}{P_0(x)+qM(x)W(x-\hbar)}\nn\\
&=&-\frac{1}{P_0(x)-\frac{q M(x)}{P_0(x-\hbar)-\frac{q M(x-\hbar)}{P_0(x-2\hbar)-\dots}}}\,.\qquad
\eea
Equivalently, one can solve the recursion for $W(x-\hbar)$. Comparing the two results one obtains the following expression (see also \cite{Poghosyan:2020zzg,Cipriani:2025ikx})
\bea
\frac{M(a)}{P(a{-}1){-}\frac{M(a{-}1)}{
   P(a{-}2){-}\dots}}{+}\frac{M(a{+}1)}{P(a{+}1){-}
   \frac{M(a{+}2)}{P(a{+}2){-}\dots}}{-}P(a){=}0\nn\\
\eea
where $x=a$ and $\hbar=1$. The previous relation can be solved perturbatively in $q$ to find the fundamental cycle $a$. We have
\bea\label{acycle}
a_{\ell=0}&=&\frac{1}{2}-\frac{\sqrt{5}}{2}q-\frac{21}{8\sqrt{5}}q^2-\frac{274}{45\sqrt{5}}q^3-\frac{471527}{28800\sqrt{5}}q^4\nn\\
&+&O\left(q^5\right)\,,\nn\\
a_{\ell=1}&=&1-\frac{\sqrt{35}}{8}q-\frac{57}{128}\sqrt{\frac{7}{5}}q^2-\frac{3388807}{737280\sqrt{35}}q^3\nn\\
&-&\frac{497698171}{58982400\sqrt{35}}q^4+O\left(q^5\right)\,,\nn\\
a_{\ell=2}&=&\frac{3}{2}-\frac{1}{2}q-\frac{7}{40}q^2-\frac{67}{450}q^3-\frac{33853}{201600}q^4\nonumber\\
&+& O\left(q^5\right)\,.
\eea
For generic $\ell$ one finds
\begin{widetext}
\bea
\label{a_l_generico}
a&=&\frac{\ell+1}{2}-\frac{3q}{2(\ell+1)}+\frac{48+170\ell-55\ell^2-140\ell^3-35\ell^4}{8\ell(\ell-1)(\ell+1)^3(\ell+2)(\ell+3)}q^2\nn\\
&-&\frac{3 \left(77 \ell^8+616
   \ell^7+1526 \ell^6+532 \ell^5-2527
   \ell^4-2156 \ell^3+924 \ell^2+1008
   \ell+288\right) q^3}{8 (\ell-1)^2
   \ell^2 (\ell+1)^5 (\ell+2)^2 (\ell+3)^2}\nn\\
   &-&\frac{q^4}{128 (\ell-3) (\ell-2) (\ell-1)^3 \ell^3
   (\ell+1)^7 (\ell+2)^3 (\ell+3)^3 (\ell+4)
   (\ell+5)}\Big[3 \left(10725 \ell^{16}+171600
   \ell^{15}+898040 \ell^{14}\right.\nn\\
   &+&\left.560560
   \ell^{13}-10797006
   \ell^{12}-33788392 \ell^{11}+622864
   \ell^{10}+150160560
   \ell^9+180773857 \ell^8-147304696
   \ell^7\right.\nn\\
   &-&\left.356718768 \ell^6-19871120
   \ell^5+250983344 \ell^4+102763968
   \ell^3-39714816 \ell^2-47646720
   \ell-11059200\right)\Big]\nn\\
   &+&O\left(q^5\right)\,,
\eea
with $\nu=2a-1$ in five-dimensions.
\end{widetext}

\section{Mano-Suzuki-Takasugi like approach}

In this section we will outline the construction of the MST-type in and up solutions for the RCHE discussed above. Noticeably, this generalization from the simple CHE is new in the literature  and constitutes an original contribution of the present study. This extension of the MST technique to the CHE paves the way to the application of this kind of treatment to other solutions, such as fuzzballs, including the D1-D5 circular-profile fuzzball and JMaRT \cite{Bianchi:2022qph,Bianchi:2023rlt}.

\subsection{MST-like In solution}
Using the effective potential \eqref{Q20} and using the dictionary \eqref{dict}, we have
\beq
\psi''(y)+Q_{0,2}(y)\psi(y)=0\,,
\eeq
with
\beq
Q_{0,2}=\frac{1-Ly(1+y)+4q(1+y)^3}{4y^2(1+y)^2}\,.
\eeq
Let us define the new radial wave function
\beq
\psi(y)=y^{\frac{1}{2}-i\sqrt{q}}\sqrt{1+y}R_{\rm in}(y)\,,
\eeq
where $y^{\frac{1}{2}-i\sqrt{q}}$ ensures the ingoing boundary condition at the horizon $y=0$. The differential equation becomes
\bea\label{eq:in}
&&y(1+y)R_{\rm in}''(y)+\Big[1-2i\sqrt{q}+2(1-i\sqrt{q})y\Big]R'_{\rm in}(y)\nn\\
&&+\Big[\frac{1}{4}\left(4q-4i\sqrt{q}-\ell(\ell+2)\right)+q y\Big]R_{\rm in}(y)=0\,.
\eea
Let us rewrite the differential equation in the following way
\bea
\label{eq_R_in}
&&y(1+y)R''_{\rm in}(y)+\Big[1-2i\sqrt{q}+2(1-i\sqrt{q})y\Big]R'_{\rm in}(y)\nn\\
&&-\frac{1}{4}\left(\nu^2+2\nu+4i\sqrt{q}+4q\right)R_{\rm in}(y)\nn\\
&=&\Big[\frac{1}{4}\left(\ell^2+2\ell-\nu^2-2\nu\right)-2q-q y\Big]R_{\rm in}(y)=0\,,
\eea
where we introduced the renormalized angular momentum $\nu$, whose leading behavior is given by $\nu=\ell+O(q)$. Notice that the right-hand side of the previous equation can be treated as a source term which, being proportional to the gauge coupling $q$, is assumed to be small. The regular solution of the left hand side is 
\bea\label{2f1in}
F_\nu(y){=}{}_2F_1\left({-}i\sqrt{q}{-}\frac{\nu}{2},1{-}i\sqrt{q}{+}\frac{\nu}{2},1{-}2i\sqrt{q},{-}y\right)\,.
\eea
Replacing $\nu\to \nu+n$ in \eqref{2f1in}, we can rewrite $y R_{\rm in }(y)$ appearing in the source term as
\bea
&&y F_{\nu{+}n}(y){=}A^{\rm in}_+ F_{n{+}2{+}\nu}(y){+}A^{\rm in }_0 F_{n{+}\nu}(y){+}A^{\rm in}_- F_{n{-}2{+}\nu}(y)\,,\nn\\
&&A^{\rm in}_+=\frac{(2+n-2i\sqrt{q}+\nu)^2}{4(1+n+\nu)(2+n+\nu)}\,,\nn\\
&&A^{\rm in}_0=-\frac{1}{2}+\frac{2q}{(n+\nu)(2+n+\nu)}\,,\nn\\
&&A^{\rm in}_-=\frac{(n+2i\sqrt{q}+\nu)^2}{4(n+\nu)(1+n+\nu)}\,.
\eea
We can then write the solution of the complete differential equation as
\beq\label{inexp}
R_{\rm in}(y)=\sum_{n=-\infty}^\infty a_n F_{\nu+n}(y)\,.
\eeq
Using the previous \lq\lq shift relation" we can write the whole differential equation as 
\bea
&&\frac{1}{4}n(n+2+2\nu)F_{n+\nu}(y)-\Big[-q(A^{\rm in}_+ F_{n+2+\nu}(y)\nn\\
&+&A^{\rm in }_0 F_{n+\nu}(y)+A^{\rm in}_- F_{n-2+\nu}(y))\nn\\
&+&\frac{1}{4}(\ell^2+2\ell-\nu^2-\nu-8q)F_{n+\nu}(y)\Big]=0\,,
\eea
which can be symbolically written as a five terms recursion of the form
\bea\label{5term}
&&\alpha_n a_{n+2}+\gamma_n a_n+\epsilon_n a_{n-2}=0\,,\nn\\
&&\alpha_n=\frac{q(2+n+2i\sqrt{q}+\nu)^2}{4(2+n+\nu)(3+n+\nu)}\,,\nn\\
&&\gamma_n=\frac{1}{4}\Big[-\ell(\ell+2)+n^2+6q+2n(1+\nu)\nn\\
&&+\nu(2+\nu)+\frac{8q^2}{(n+\nu)(2+n+\nu)}\Big]\,,\nn\\
&&\epsilon_n=\frac{q(n-2i\sqrt{q}+\nu)^2}{4(n+\nu-1)(n+\nu)}\,.
\eea

Recall that the Gauss hypergeometric function in Eq.~\eqref{2f1in} solves an ODE with regular singularities at $0,-1,\infty$. However, Eq.~\eqref{eq:in} is regular singular at $y=0,-1$ and irregular at $y=\infty$. Hence, we can study the asymptotic behaviour of the solutions at $\infty$ within the framework of Borel-Laplace summability and resurgence \cite{Ecalle:1981,Mitschi:2016fxp}. This approach has the advantage of working direcly with the differential equation rather than with the infinite series in Eq.~\eqref{inexp}. 

\begin{figure}
\includegraphics[scale=0.4]{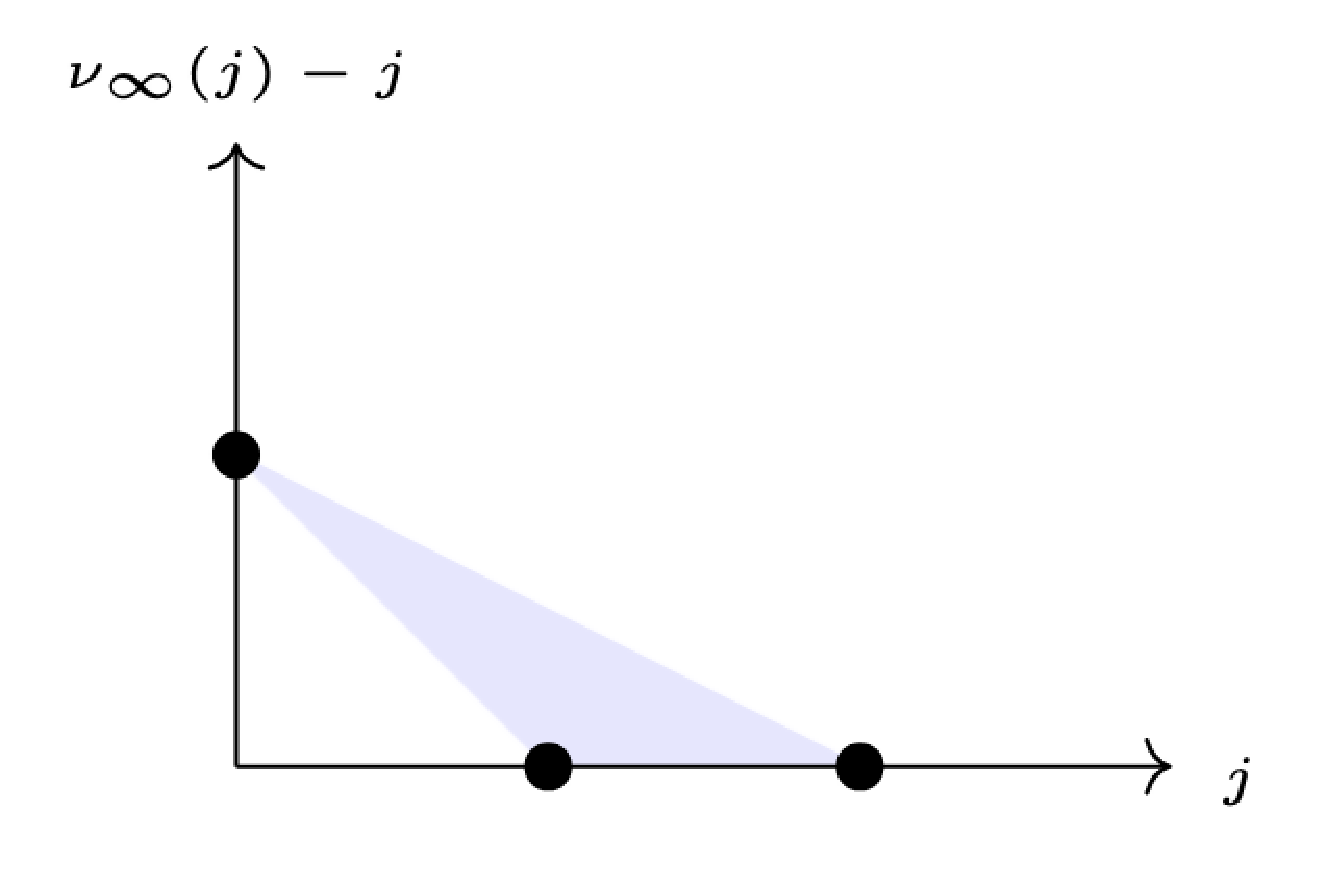}
\caption{\label{fig:NP-in} The Newton's polygon of Eq. \eqref{eq:in} at $y =\infty$, assuming $q\neq 0,-1$. Following the convention of~\cite[Sec.~3.3.3.1]{diver-ii}, the points of the polygon are $(j,\nu_\infty(j)-j)$ for $j=0,1,2$, where $\nu_\infty(j)$ is the degree of the polynomial coefficient of the $j$-th derivative in Eq.~\eqref{eq:in}. In particular, the points are located at $(2,0), (1,0)$ and $(0,1)$. The levels correspond to the absolute value of (negative) slopes of the polygon, namely $1/2$ and $1$.}
\end{figure}

First, assuming $q\neq 0,-1$,  from the Newton's polygon, i.e., a tool to reconstruct the behaviour of a differential equation close to its singular points, 
(in Fig.~\ref{fig:NP-in}) we deduce that the irregular singularity are of level $1/2$ and $1$, hence the solution has an asymptotic behaviour of the form 
\bea
\label{eq:Rin-asym}
&R_{\rm in }(y)\sim e^{\pm \lambda_1 \sqrt{y}}\sum_{n=0}^\infty a^{(1)}_{\pm,n} y^{-n/2} \nn\\
&\qquad + e^{\pm \lambda_2 y}\sum_{n=0}^\infty a^{(2)}_{n} y^{-n}\,, \quad y\to \infty\,,
\eea
where $\lambda_1=\sqrt{-q}$, $\lambda_2=\frac{q}{2(1-i\sqrt{q})}$ and the coefficients $a^{(1)}_{\pm,n}$ (resp. $a^{(2)}_{n}$) grow as $(n!)^2$ (resp. $n!$).  Then, the theory of multi-summability guarantees the existence of an analytic solution in the form of generalized Borel-Laplace sum (See e.g., Ref.~\cite{Ramis:1993}, also for notational details).
 The domains of analyticity of such solutions and their exact analytic expressions will require an {\it ad-hoc} analysis, which is beyond the scope of the present work.

\subsection{MST-like Up solution}

In \eqref{eqR} let us introduce the new variable 
\beq
r=\frac{2\sqrt{z}}{\omega}\,,
\eeq
so that the equation becomes
\bea
&&R''(z)+\left(\frac{1}{z}+\frac{1}{z-q}\right)R'(z)\nn\\
&&\Big[\frac{{\mathcal L}}{4qz}+\frac{q}{(z-q)^2}+\frac{4q-{\mathcal L}}{4q(z-q)}\Big]R(z)=0\,.
\eea
Let us rescale the wavefunction as
\bea
R(z)=K(z)(z-q)^{-i\sqrt{q}}z^{-1+i\sqrt{q}}\,.
\eea
The differential equation becomes
\bea\label{eq:up}
&&0=z(z-q)K''(z)+(1-2i\sqrt{q})qK'(z)\nn\\
&&+\Big[z{+}q{-}\frac{{\mathcal L}}{4}{+}\frac{q^2{-}q{+}2iq^{3/2}}{z}\Big]K(z)\,.
\eea
Using the same ideas of the in solution, let us split the differential equation as follows
\bea
\label{eqK}
&&z^2K''(z)+\Big[z-\frac{\nu^2+2\nu}{4}\Big]K(z)\nn\\
&&=qzK''(z)+(-1+2i\sqrt{q})qK'(z)\nn\\
&&{+}\Big[\frac{1}{4}(\ell^2{+}2\ell{-}\nu^2{-}2\nu{-}4q){+}\frac{q{-}2iq^{3/2}{-}q^2}{z}\Big]K(z)\,,\nn\\
\eea
where we recall that the quantity $\ell^2{+}2\ell{-}\nu^2{-}2\nu\sim O(q)$.

The left hand side is solved in terms of Bessel functions. The outgoing wave function at infinity reads
\beq\label{upsol}
f_{\nu+n}(z){=}i^n\sqrt{z}\Big[e^{i\pi(n{+}\nu)}J_{{-}1{-}n{-}\nu}(2\sqrt{z}){+}J_{1{+}n{+}\nu}(2\sqrt{z})\Big]\,,
\eeq
with $f_{\nu+n}(z)$ of Eq. \eqref{upsol} for $n=0$ satisfies the left-hand-side of Eq. \eqref{eqK}.

Consequently, the solution of the whole differential equation reads
\beq\label{upexp}
K(z)=\sum_{n=-\infty}^\infty a_n f_{\nu+n}(z)\,.
\eeq
In the previous expression the presence of $i^n$ is necessary in order to obtain a five term recursion as in \eqref{5term}.

The following identities hold
\beq
f'_{n+\nu}(z)=A^{\rm up}_+ f_{n+2+\nu}(z)+A^{\rm up}_0 f_{n+\nu}(z)+A^{\rm up}_- f_{n-2+\nu}(z)\,,
\eeq
with
\bea
&&A^{\rm up}_+=\frac{n+\nu}{2(1+n+\nu)(2+n+\nu)}\,,\nn\\
&&A^{\rm up}_0=\frac{2}{(n+\nu)(n+\nu+2)}\,,\nn\\
&&A^{\rm up}_-=-\frac{2+n+\nu}{2(n+\nu)(1+n+\nu)}\,,
\eea
and
\beq
\frac{f_{n+\nu}(z)}{z}=B^{\rm up}_+ f_{n+2+\nu}(z)+B^{\rm up}_0 f_{n+\nu}(z)+B^{\rm up}_- f_{n-2+\nu}(z)\,,
\eeq
with
\bea
&&B^{\rm up}_+=\frac{-1}{(1+n+\nu)(2+n+\nu)}\,,\nn\\
&&B^{\rm up}_0=\frac{2}{(n+\nu)(2+n+\nu)}\,,\nn\\
&&B^{\rm up}_-=-\frac{1}{(n+\nu)(1+n+\nu)}\,.
\eea
Consequently,  the differential equation can be rewritten as
\bea
&&\frac{1}{4}n(2+n+2\nu)f_{n+\nu}(z)\nn\\
&&-\Big[\frac{1}{4}\Big(\ell^2+2\ell-\nu^2-2\nu-8q\Big)f_{n+\nu}(z)\nn\\
&&+\frac{1}{4}\Big(4q+2nq+n^2q-8iq^{3/2}-4q^2+2q\nu+2nq\nu\nn\\
&&{+}q\nu^2\Big)(B^{\rm up}_+ f_{n{+}2{+}\nu}(z){+}B^{\rm up}_0 f_{n{+}\nu}(z){+}B^{\rm up}_- f_{n{-}2+\nu}(z))\nn\\
&&+i(i+2\sqrt{q})q(A^{\rm up}_+ f_{n+2+\nu}(z)+A^{\rm up}_0 f_{n+\nu}(z)\nn\\
&&+A^{\rm up}_- f_{n-2+\nu}(z))\Big]=0\,.
\eea
The choice of $i^n$ appearing in \eqref{upsol} ensures that the previous expression can be rewritten exactly as \eqref{5term}.

\subsection{Recursion relation}

The five-terms recursion can be treated as two independent three terms recursion in the following way
\bea
\alpha^{(e)}_na_{2n+2}+\beta^{(e)}_na_{2n}+\gamma^{(e)}_n a_{2n-2}=0\,,\nn\\
\alpha^{(o)}_na_{2n+1}+\beta^{(o)}_na_{2n-1}+\gamma^{(o)}_n a_{2n-3}=0\,,
\eea
where
\bea
&&\alpha^{(e)}_n=\alpha_{2n},\quad \beta^{(e)}_n=\gamma_{2n},\quad \gamma^{(e)}_n=\epsilon_{2n}\,,\nn\\
&&\alpha^{(o)}_n=\alpha_{2n+1},\quad \beta^{(0)}_n=\gamma_{2n+1},\quad \gamma^{(o)}_n=\epsilon_{2n+1}\,.\nn\\
\eea
Performing the shift $n\to n+1$ in the odd recursion relation, we obtain
\bea
\alpha^{(o)}_{n+1}a_{2n+3}+\beta^{(o)}_{n+1}a_{2n+1}+\gamma^{(o)}_{n+1} a_{2n-1}=0\,,
\eea
where is easy to notice that
\bea
&&\alpha^{(o)}_{n+1}=\alpha^{(e)}_{n},\quad \beta^{(o)}_{n+1}=\beta^{(e)}_{n},\quad \gamma^{(o)}_{n+1}=\gamma^{(e)}_{n}\,.\qquad
\eea
Therefore,  both the three terms recursions have the same coefficients, and from now on we will avoid the even-odd extra labels \lq\lq e" and \lq\lq o." The corresponding continuous fraction is given by
\bea
\beta _0-\frac{\alpha _{-1} \gamma _0}{\beta _{-1}-\frac{\alpha _{-2} \gamma
   _{-1}}{\beta _{-2}-\dots}}-\frac{\alpha _0 \gamma _1}{\beta _1-\frac{\alpha _1
   \gamma _2}{\beta _2-\dots}}=0\,,
\eea
and it can be solved perturbatively in $\nu$, considering $q$ a small quantity. For example, the result for $\ell=2$ is
\bea
&&\nu_{\ell=2}=2-\frac{1}{4} M^2 \omega ^2-\frac{7}{320} M^4 \omega ^4-\frac{67 M^6 \omega ^6}{14400}\nn\\
&&-\frac{33853 M^8 \omega
   ^8}{25804800}+O(M^{10}\omega^{10})\,.
\eea
Comparing with the fundamental SW cycle \eqref{acycle} we have 
\beq
\nu=2a-1=2\left(a-\frac12 \right)\,,
\eeq
differently to the relation in four dimensions $\nu=a-1/2$, suggesting, for example,  
a general relation of the type
\beq
\nu=(d-3)\left(a-\frac12 \right)\,.
\eeq
This \lq\lq guessed" behaviour, however, requires additional studies and checks.
Let us observe that, in the reconstruction of the \lq in\rq{} \eqref{inexp} and \lq up\rq{} \eqref{upexp} solutions, no contribution from the odd part of the recursion is allowed in order to satisfy the radial differential equation \eqref{eqR}. Therefore, we must set the coefficient $a_{1}=0$.

\begin{figure}
\includegraphics[scale=0.4]{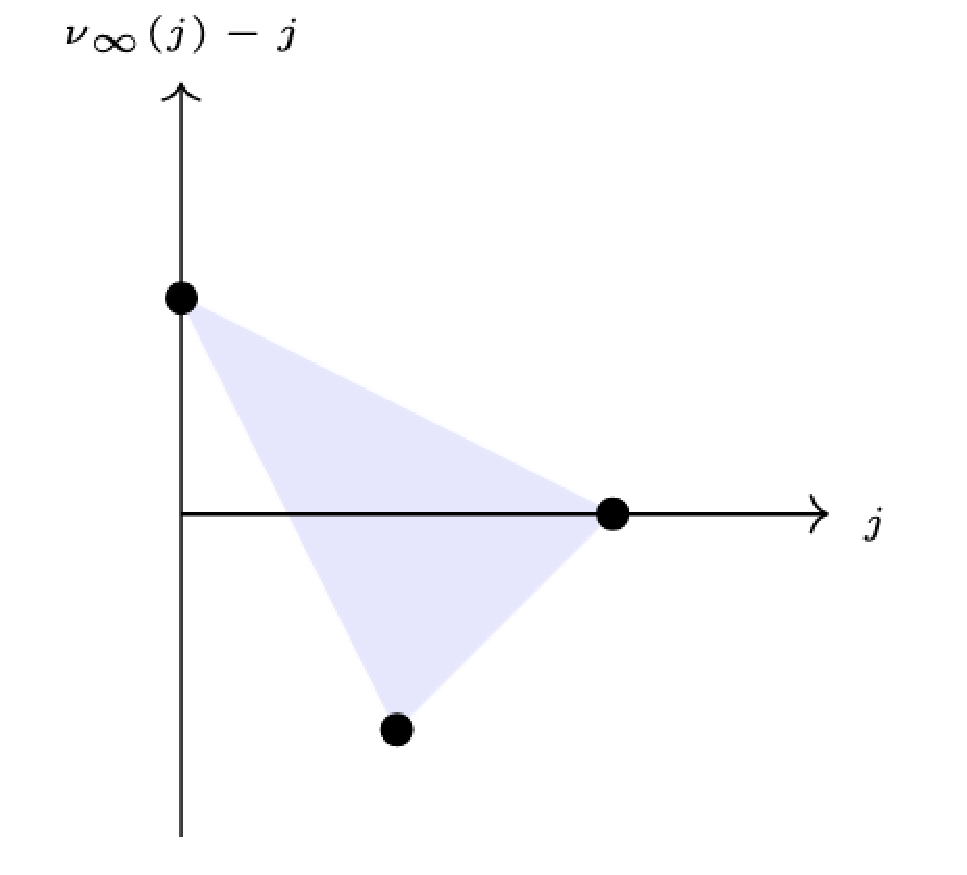}
\caption{\label{fig:NP-up}The Newton' polygon of Eq.~\eqref{eq:up} at $z=\infty$, assuming $q\neq 0,-1/4$. Following the convention of~\cite[Sec.~3.3.3.1]{diver-ii}, we define the points of the polygon as $(j,\nu_\infty(j)-j)$ for $j=0,1,2$, where $\nu_\infty(j)$ is the degree of the polynomial coefficient of the $j$-th derivative in Eq.~\eqref{eq:up}. In particular, the points are located at $(2,0), (1,-1)$ and $(0,1)$. The levels are given by the absolute values of (negative) slopes of the polygon, namely $1/2$ and $2$.}
\end{figure}

Moreover, let us notice that Eq.~\eqref{eq:up} has regular singularities at $z=0,q$ and an irregular singularity at $z=\infty$. 
The latter can be studied by looking at the Newton's polygon (in Fig.~\ref{fig:NP-up}), from which we deduce that if $q\neq 0,-1/4$ the irregular singularity is of multiple levels $1/2$ and $2$. Hence, the solution has an asymptotic behaviour of the form 
\bea
K(z)\sim &&e^{\pm \lambda_1 \sqrt{z}}\sum_{n=0}^\infty a^{(1)}_{\pm,n} z^{-n/2}\nn \\
&&+ e^{z^2/\lambda_2}\sum_{n=0}^\infty a^{(2)}_{n} z^{-2n}\,, \quad z\to \infty\,,
\eea
where $\lambda_1=i$, $\lambda_2=(1-2i\sqrt{q})q$ and the coefficients $a^{(1)}_{\pm,n}$ (resp. $a^{(2)}_{n}$) grow as $(n!)^2$ (resp. $(n!)^{1/2}$). Also in this case, the theory of multi-summability guarantees the existence of an analytic solution $K(z)$ in the form of generalized Borel-Laplace sum. If $(1-2i\sqrt{q})q= 0$, then the irregular singularity is of single level $1/2$ and  the solution $K$ will behave asymptotically as $\exp(\pm\lambda_1\sqrt{z})$.

We summarize in Table \ref{tab1:leq2} the explicit in and up solutions of MST type for $\ell=2$.

\begin{table*}
\caption{\label{tab1:leq2} Explicit construction of the in and and up solutions for $\ell=2$ with a MST like procedure.  }
\begin{ruledtabular}
\begin{tabular}{ll}
$R_{\rm in}^{\ell=2}$ & $\frac{2 r^2}{M^2} +\eta^2\left[ -\frac{r^4 \omega ^2}{8
   M^2}+\frac{2 i r^2 \omega
   }{M}-1 \right] $\\
&$+\eta^4 \left[\frac{1}{960} \omega
   \left(\frac{3 r^6 \omega
   ^3}{M^2}-\frac{120 i r^4 \omega
   ^2}{M}+480 r^2 \omega  \log
   \left(\frac{M}{r}\right)-960 i
   M-40 \left(39+2 \pi ^2\right)
   r^2 \omega \right)  \right]$\\
&$+\eta^6 \left[ \frac{\omega ^2 \left(960
   \left(30+\pi ^2\right) M^4-2880
   i M^3 r^2 \omega  \left(4 \zeta
   (3)+7+\pi ^2\right)+24
   \left(93+5 \pi ^2\right) M^2
   r^4 \omega ^2-720 M^2 \left(8
   M^2-16 i M r^2 \omega +r^4
   \omega ^2\right) \log
   \left(\frac{M}{r}\right)+72 i M
   r^6 \omega ^3+r^8 \left(-\omega
   ^4\right)\right)}{23040 M^2} \right]$\\
&$+\eta^8 \Big[ 
\frac{\omega}{12902400}  \Big(\frac{1075200
   i M^5}{r^4}+\frac{1612800 M^4
   \omega }{r^2}+806400 i M^3
   \omega ^2 \left(4 \zeta
   (3)+14+\pi ^2\right)+\frac{5
   r^{10} \omega ^7}{M^2}+806400
   M^2 r^2 \omega ^3 \log
   ^2\left(\frac{M}{r}\right)$\\
&$+896
   M^2 r^2 \omega ^3 \left(9000
   \zeta (3)+3862+1650 \pi ^2+95
   \pi ^4\right)+3360 \omega ^2
   \left(-960 i M^3-16 \left(87+5
   \pi ^2\right) M^2 r^2 \omega    -120 i M r^4 \omega ^2+3 r^6
   \omega ^3\right) \log
   \left(\frac{M}{r}\right)$\\
&$-\frac{ 
   560 i r^8 \omega ^6}{M}+20160 i
   M r^4 \omega ^4 \left(20 \zeta
   (3)+32+5 \pi ^2\right)-56
   \left(529+30 \pi ^2\right) r^6
   \omega ^5\Big) \Big]$\\
$R_{\rm up}^{\ell=2}$ & $ \frac{8 M^2}{r^4 \omega ^2}+\eta^2 \left[ \frac{M^2 \left(\frac{8
   M^2}{\omega
   ^2}+r^4\right)}{r^6}
 \right]$\\
&$+\eta^4 \left[ \frac{36 M^6}{5 r^8 \omega
   ^2}+\frac{2 M^4 \log
   \left(\frac{1}{2} e^{\gamma } r
   \omega \right)}{r^4}-\frac{i
   (10 \pi -9 i) M^4}{5
   r^4}+\frac{M^2 \omega ^2}{8} \right]$\\
&$+\eta^6 \Big[
\frac{32 M^8}{5 r^{10} \omega
   ^2}-\frac{i (2 \pi -3 i)
   M^6}{r^6}+\frac{i M^5 \omega
   }{4 r^4}+\frac{(-13-20 i \pi )
   M^4 \omega ^2}{80
   r^2}+\frac{1}{576} (11+6 i \pi
   ) M^2 r^2 \omega
   ^4+\left(\frac{2
   M^6}{r^6}+\frac{M^4 \omega
   ^2}{4 r^2}-\frac{1}{48} M^2 r^2
   \omega ^4\right) \log
   \left(\frac{1}{2} e^{\gamma } r
   \omega \right)
\Big]$\\
& $+\eta^8 \Big[
\frac{40 M^{10}}{7 r^{12} \omega
   ^2}-\frac{i (180 \pi -341 i)
   M^8}{100 r^8}+\frac{i M^7
   \omega }{4 r^6}+\frac{M^6
   \omega ^2 \log
   ^2\left(\frac{1}{2} e^{\gamma }
   r \omega \right)}{4
   r^4}-\frac{\left(713-660 i \pi
   +700 \pi ^2\right) M^6 \omega
   ^2}{2400 r^4}+\frac{i M^5
   \omega ^3}{32
   r^2}+\frac{(19-105 i \pi ) M^4
   \omega ^4}{2880}$\\
&$+\frac{(-37-12
   i \pi ) M^2 r^4 \omega
   ^6}{18432}+\left(\frac{9 M^8}{5
   r^8}-\frac{i (20 \pi -11 i) M^6
   \omega ^2}{40 r^4}+\frac{M^4
   \omega ^4}{24}+\frac{1}{768}
   M^2 r^4 \omega ^6\right) \log
   \left(\frac{1}{2} e^{\gamma } r
   \omega \right)
\Big]$\\
\end{tabular}
\end{ruledtabular}
\end{table*}

\section{Application: Circular Orbits and Energy Loss}

\subsection{Radiated energy and angular momentum}

In the following, we will consider the loss of energy and angular momentum by emission of (massless) scalar waves
when the scalar field is sourced by a scalar charge $q$ in a circular equatorial motion
\beq\label{nonhomo}
\Box \Psi \equiv \frac{1}{\sqrt{-g}} \partial_\mu \left(\sqrt{-g}g^{\mu\nu}\partial_\nu \Psi \right)=-4\pi \rho\,,
\eeq 
with $\sqrt{-g}=r^3 \sin \theta \cos \theta$ and
\beq\label{sourverho}
\rho(x)=q\int \frac{d\tau}{\sqrt{-g}}\delta^{(5)}(x-x_p(\tau))\,.
\eeq
Here we assume for the source
\bea
t&=&\Gamma \tau,\quad \phi=\Omega t\,,\quad r=r_0\,,\nonumber\\ 
\theta &=& \theta_0=\frac{\pi}{2}\,,\quad \psi=\psi_0=0\,, 
\eea
with
\beq
\Gamma=\frac{1}{\sqrt{1-\frac{2 M^2}{r_0^2}} }\,,\qquad
\Omega=\frac{  M }{r_0^2}\,,
\eeq
so that
\bea
\rho(x)&=& \frac{q}{\Gamma r_0^3}\delta (r-r_0)\frac{\delta (\phi-\Omega t)\delta(\theta-\frac{\pi}{2})\delta(\psi)}{\sin\theta \cos\theta}\nonumber\\
&=& \frac{q}{\Gamma r_0^3}\delta (r-r_0)\delta_{S^3}(\theta,\phi,\psi;\frac{\pi}{2},\Omega t,0)\,,
\eea
where we recall that the distribution
\beq
\delta_{S^3}(\chi,\chi')=\frac{\delta(\theta-\theta')\delta(\phi-\phi')\delta(\psi-\psi')}{\sin\theta \cos\theta}
\eeq
with $\chi=(\theta,\phi,\psi)$ and $\chi'=(\theta',\phi',\psi')$
is well defined when integrated on the the $S_3$, with corresponding measure  $d\Omega_3=\sin\theta \cos\theta d\theta d\phi d\psi$.

The formulae for the radiated energy and angular momentum can be derived from the associated energy-momentum tensor which  
for a massless complex scalar field is given by
\beq
8\pi T^{\rm scal}_{\mu\nu}=\partial_\mu\Psi^*\partial_\nu\Psi+ \partial_\mu\Psi \partial_\nu\Psi^* -  g_{\mu\nu} \partial_\lambda \Psi^* \partial^\lambda\Psi \,,
\eeq
so that
\beq\label{fluxinf}
\frac{d^2E}{dtd\Omega_3}=\lim_{r\to \infty} (r^3 T^{\rm scal}{}^r{}_t)\,,
\eeq
where we have $r^3$ instead of $r^2$ as in the $d=4$ case.
We find
\beq
8\pi T^{\rm scal}_{rt}=\Psi^*_{,r}\Psi_{,t}+\Psi_{,r}\Psi^*_{,t} \,,
\eeq
implying
\bea
8\pi T^{\rm scal}{}^r{}_{t}&=& g^{rr}(\Psi^*_{,r}\Psi_{,t}+\Psi_{,r}\Psi^*_{,t})\nonumber\\
&=&\left(1-\frac{M^2}{r^2} \right) \Psi^*_{,r}\Psi_{,t}+{\rm c.c.}\,.\qquad 
\eea
Let us write
\beq
\Psi= \sum_{\ell m_\phi m_\psi } Y_{\ell m_\phi m_\psi}(\theta,\phi,\psi) \int \frac{d\omega}{2\pi}e^{-i\omega t}R_{\ell m_\phi m_\psi \omega}(r)\,,
\eeq
where
\begin{widetext}
\bea
Y_{\ell m_\phi m_\psi}(\theta,\phi,\psi)&=&e^{im_\phi \phi}e^{im_\psi \psi}S_{\ell m_\phi m_\psi}(\theta)\nn\\
\eea
where $S_{\ell m_\phi m_\psi}(\theta)$ are related to the Jacobi polynomials. 
Introducing the notation
\beq
\ell_{_{(\pm,\pm)}}=\frac12 (\ell \pm | m_{\phi } |\pm | m_{\psi }|)\,,
\eeq
one finds 
\bea
S_{\ell m_\phi m_\psi}(\theta)&=&\mathcal{N}_{\ell m_\phi m_\psi}\cos ^{ | m_{\psi } |}(\theta ) \sin ^{ | m_{\phi} | }(\theta ) \,
   _2F_1\left(- \ell_{_{(-,-)}}, \ell_{_{(+,+)}}+1; |m_{\psi } | +1;\cos^2(\theta )\right)\,,
\eea
and
\bea
\mathcal{N}_{\ell m_\phi m_\psi}&=&\frac{i \sqrt{2(\ell +1)}}{| m_{\psi}| !} \frac{
   \sqrt{\ell_{_{(-,+)}}!}
   \sqrt{\ell_{_{(+,+)}}!}}{
   \sqrt{\ell_{_{(-,-)}}!}
   \sqrt{\ell_{_{(+,-)}}!}}\,.
\eea

Finally, let us recall
\begin{enumerate}
\item The orthogonality relation of the spherical harmonics on $S^3$ 
\beq\label{ortho}
\int_{S^3} d\Omega_3 Y_{\ell m_\phi m_\psi}Y^*_{\ell' m_\phi' m_\psi'}=\delta_{\ell \ell'}\delta_{m_\phi m_\phi'}\delta_{m_\psi m_\psi'}\,.
\eeq
\item The completeness relation for spherical harmonics on $S^3$ 
\bea
\delta_{S_3}(\chi,\chi')=\sum_{\ell m_\phi m_\psi}Y_{\ell m_\phi m_\psi}(\chi)Y^*_{\ell m_\phi m_\psi}(\chi')\,,
\eea
where $\chi=(\theta,\phi,\psi)$ is a commonly used notation for all the angular variables.
\end{enumerate}
The non homogeneous equation \eqref{nonhomo} in presence of the source \eqref{sourverho} can be made explicit as follows
\bea
\Box \Psi=\sum_{\ell m_\phi m_\psi}\int \frac{d\omega}{2\pi}e^{-i\omega t}Y_{\ell m_\phi m_\psi}(\theta,\phi,\psi)\mathcal{L}_r R_{\ell m_\phi m_\psi \omega}(r)=-4\pi \int\frac{d\omega}{2\pi}e^{-i\omega t}\mathcal{S}_{\ell m_\phi m_\psi\omega}(r,\chi)\,,
\eea
where
\bea
\mathcal{S}_{\ell m_\phi m_\psi\omega}(r,\chi)&=&\int dt e^{i\omega t}\rho\nn\\
\mathcal{L}_r&=&\left(1-\frac{M^2}{r^2}\right)\frac{d^2}{dr^2}+\frac{3r^2-M^2}{r^3}\frac{d}{dr}+\Big[\frac{\omega^2 r^2}{r^2-M^2}-\frac{\ell(\ell+2)}{r^2}\Big]\,.
\eea
Consequently,
\beq
\mathcal{L}_r R_{\ell m_\phi m_\psi \omega}(r)=-\frac{4\pi q}{r_0^3}\int dt e^{i\omega t}\int d\tau \delta(t-\Gamma\tau)\delta(r-r_0)Y^*_{\ell m_\phi m_\psi}\left(\frac{\pi}{2},\Gamma \Omega\tau,0\right)
\eeq
where we recall our choice $\psi=0$. Furthermore 
\beq
Y^*_{\ell m_\phi m_\psi}\left(\frac{\pi}{2},\Omega t,0\right)=-i\sqrt{2(\ell+1)}\delta_{m_\psi,0}e^{-im_\phi \Omega t }\,.
\eeq
Therefore,
\beq\label{720}
\mathcal{L}_r R_{\ell m_\phi m_\psi \omega}(r)=i\frac{8\pi^2 q}{r_0^4}\sqrt{r_0^2-2M^2}\sqrt{2(\ell+1)}\delta(r-r_0)\delta_{m_\psi,0}\delta(\omega-m_\phi \Omega)\,.
\eeq
The radiated energy by emission of massless scalar waves at infinity \eqref{fluxinf} can be made more explicit as
    \bea
r^3 {T^{\rm scal}{}^r}_t&=&\frac{r(r^2-M^2)}{8\pi}\sum_{\ell m_\phi m_\psi}\sum_{\ell' m_\phi' m_\psi'}\int\frac{d\omega}{2\pi}\int\frac{d\omega'}{2\pi}Y_{\ell m_\phi m_\psi}(\theta,\phi,\psi)Y^*_{\ell' m_\phi' m_\psi'}(\theta,\phi,\psi)\nn\\
&\times&\Big[(-i\omega)e^{-i(\omega-\omega')t}R_{\ell m_\phi m_\psi\omega}\frac{d}{dr}R^*_{\ell'm_\phi'm_\psi'\omega'}+(i\omega')e^{i(\omega-\omega')t}R^*_{\ell'm_\phi'm_\psi'\omega'}\frac{d}{dr}R_{\ell m_\phi m_\psi\omega}\Big]\,.
\eea
Using the orthogonality relation \eqref{ortho} we can rewrite the energy flux as
\bea\label{dedt}
\frac{dE}{dt}&=&\lim_{r\to \infty}\int \sin \theta \cos \theta d\theta d\phi d\psi\,\,  r^3 T^{\rm scal}{}^r{}_{t}\nonumber\\
&=&\lim_{r\to \infty} \frac{r(r^2-M^2)}{8\pi} \sum_{\ell m_\phi m_\psi}\int \frac{d\omega}{2\pi} \frac{d\omega'}{2\pi}\Big[(-i\omega) e^{-i(\omega-\omega')t} R_{\ell m_\phi m_\psi \omega}(r)\frac{d}{dr}R^*_{\ell m_\phi m_\psi \omega'}(r)\nonumber\\
&+&  (i\omega')e^{i(\omega-\omega')t} R^*_{\ell m_\phi m_\psi\omega'}(r)\frac{d}{dr}R_{\ell m_\phi m_\psi \omega }(r)\Big]\,.
\eea
From now on we will drop the dependence on $m_\psi$ of the various functions, since it has to be zero for our choice of the source's orbit. 

Eq. \eqref{720} then becomes
\bea
\label{diffeqfin}
\mathcal{K}_r R_{\ell m_\phi 0 \omega}&=&\delta(r-r_0)P_{\ell m_\phi\omega}(r_0)\,,
\eea
with
\bea
P_{\ell m_\phi\omega}(r_0)&=&i\frac{8\pi^2q}{r_0^2}\frac{\sqrt{r_0^2-2M^2}}{r_0^2-M^2}\sqrt{2(\ell+1)}\delta(\omega-m_\phi \Omega)\,,\nn\\
\mathcal{K}_r&=&\frac{d^2}{dr^2}+\frac{3r^2-M^2}{r(r^2-M^2)}\frac{d}{dr}+\frac{r^2}{r^2-M^2}\Big[\frac{\omega^2r^2}{r^2-M^2}-\frac{\ell(\ell+2)}{r^2}\Big]\,.
\eea
\end{widetext}
The homogeneous part of the differential equation \eqref{diffeqfin} can be solved by using the Green function method
\beq
\mathcal{K}_r G_{\ell m_\phi \omega}(r,r')=\frac{1}{\Delta(r')}\delta(r-r')\,,
\eeq
with the Green function is defined as
\bea
G_{\ell m_\phi \omega}(r,r')&=&\frac{1}{W_{\ell m_\phi \omega}}\Big[R_{\rm in}(r)R_{\rm up}(r')\Theta(r'-r)\nn\\
&+&R_{\rm in}(r')R_{\rm up}(r)\Theta(r-r')\Big]\,.
\eea
Here the constant Wronskian is defined as
\beq
W_{\ell m_\phi \omega}=\Delta(r)\Big[R_{\rm in}(r)R'_{\rm up}(r)-R_{\rm up}(r)R'_{\rm in}(r)\Big] 
\eeq
where
\beq
\Delta(r)=r(r^2-M^2)\,.
\eeq
The particular solution can be written as
\bea
R_{\ell m_\phi \omega}(r)&=&P_{\ell m_\phi \omega}(r_0)\int dr' G_{\ell m_\phi \omega}(r,r')\Delta(r')\delta(r'-r_0)\nn\\
&=&P_{\ell m_\phi \omega}(r_0) G_{\ell m_\phi \omega}(r,r_0)\,.
\eea
In order to simplify the Fourier transforms in the frequency domain, let us introduce the following notation
\beq
R_{\ell m_\phi \omega}(r)=\mathfrak{R}_{\ell m_\phi \omega}(r)2\pi \delta(\omega-m_\phi \Omega)\,,
\eeq
where, because of the $r\to \infty$ limit, we can use here only the up-part of the $R_{\ell m_\phi \omega}(r)$ solution
\beq
\mathfrak{R}_{\ell m_\phi \omega}(r)=i\frac{4\pi q}{r_0^2}\frac{\sqrt{r_0^2-2M^2}}{r_0^2-M^2}\sqrt{2(\ell+1)}\frac{R_{\rm in}(r_0)R_{\rm up}(r)}{W_{\ell m_\phi \omega}}\,.
\eeq
The energy flux \eqref{dedt} can be rewritten as
\bea
&&\frac{dE}{dt}=\lim_{r\to +\infty}\frac{\Delta(r)}{8\pi}\sum_{\ell m_\phi}(-im_\phi \Omega)\nn\\
&\times&\Big[\mathfrak{R}_{\ell m_\phi \omega}(r)\frac{d \mathfrak{R}^*_{\ell m_\phi \omega}(r)}{dr}{-}\mathfrak{R}^*_{\ell m_\phi \omega}(r)\frac{d\mathfrak{R}_{\ell m_\phi \omega}(r)}{dr}\Big]\Bigg|_{\omega=m_\phi \Omega}\nn\\
&=&\lim_{r\to+\infty}\frac{\Delta(r)}{4\pi}\sum_{\ell m_\phi}m_\phi \Omega\,\times \nonumber\\
&& \times \mathcal{I}m\left(\mathfrak{R}_{\ell m_\phi \omega}(r)\frac{d\mathfrak{R}^*_{\ell m_\phi \omega}(r)}{dr}\right)\Bigg|_{\omega=m_\phi\Omega}\,.\nn\\
\eea

In the present investigation we limit our considerations to the $\ell=0,1,2,3$ modes,   
\bea
\frac{dE}{dt}&=&\left(\frac{dE}{dt}\right)^{\ell=0} +\left(\frac{dE}{dt}\right)^{\ell=1}+\left(\frac{dE}{dt}\right)^{\ell=2}\nonumber\\
&+& \left(\frac{dE}{dt}\right)^{\ell=3}+O(\eta^{10})\,,
\eea
and to  the NNLO approximation level we find 
\bea
\left(\frac{dE}{dt}\right)^{\ell=0}&=& O(\eta^{10})\,,\nonumber\\
\left(\frac{dE}{dt}\right)^{\ell=1}&=&-\frac{M^5\pi^2q^2\eta^4}{4r_0^{14}}\Bigg[1-\frac{2M^2\eta^2}{3r_0^2}\nn\\
&+&\frac{\eta^4M^4}{r_0^4}\Bigg(\frac{121}{576}-\frac{3\gamma}{4}-\frac{7}{8}\log\left(\frac{M}{2^\frac{8}{7}r_0}\right)\Bigg)\Bigg] \nn\\
&+& O(\eta^{10})\,,\nn\\
\left(\frac{dE}{dt}\right)^{\ell=2}&=&{-}\frac{43M^7\pi^2q^2\eta^6}{32r_0^{16}}\Bigg[1{-}\frac{515M^2\eta^2}{344r_0^2}\Bigg]{+} O(\eta^{10})\,,\nn\\
\left(\frac{dE}{dt}\right)^{\ell=3}&=&-\frac{561M^9\pi^2q^2}{128r_0^{18}}\eta^8+O(\eta^{10})\,,\nn\\
\left(\frac{dE}{dt}\right)^{\ell=4}&=&O(\eta^{10})\,. 
\eea
The sum of the various contributions written in terms of the variable
\beq
\label{u_def}
u=\frac{M^2}{r_0^2}
\eeq
yields
\bea
    \frac{dE}{dt}&=&-\frac{\pi^2q^2u^7}{4M^9}\Bigg[1+\frac{113}{24}u\nn\\
    &+&u^2\Bigg(\frac{349}{36}-\frac{3\gamma}{4}-\frac{7}{16}\log\left(\frac{u}{2^\frac{16}{7}}\right)\Bigg)\Bigg]\nn\\
    &+&O(u^{10})\,,
\eea
where we put  $\eta=1$. 

Note that in the Schwarzschild case Ref. \cite{Bini:2016egn} the expression for the scalar energy flux for circular orbits
\bea
\frac{dE_{\rm Schw}}{dt}&=&-\left(\frac{q_s}{M}\right)^2\frac13 u_s^4 \left(1-2u_s+2\pi u_s^{3/2}\right.\nonumber\\
&-& \left.10u_s^2+\frac{12}{5}\pi u_s^{5/2}\right)+O(u_s^3)\,,
\eea
with $q_s$ denotes the scalar charge of the perturbing particle in the Schwarzschild spacetime, and $u_s=M/r$ differently from the ST5d case where $u$ is defined in Eq. \eqref{u_def} coherently with the 5d case.

\section{Conclusions}

In this paper, we have investigated the five-dimensional Schwarzschild-Tangherlini solution from both a geometrical perspective and through the analysis of massless scalar perturbations.

As a starting point, we reviewed certain properties of timelike geodesics, considering both bound and unbound trajectories. In particular, we analyzed circular and hyperbolic-like orbits and study several classical aspects: geodesic deviation, parallel and Fermi-Walker transport laws, etc. In the case of unbound orbits, we computed both the radial action and the scattering angle at the probe limit, obtaining either Post-Newtonian and Post-Minkowskian expansion representations of the results (including resummation properties in terms of hypergeometric functions). We further succeeded in resumming the expanded expressions in terms of hypergeometric functions and compared them with the corresponding expressions in the more familiar four-dimensional Schwarzschild spacetime.

In addition, we extracted from the critical circular orbit regime an analytic expression for the quasinormal modes, whose imaginary part is determined by the geodesic value of the corresponding Lyapunov exponent.

The main focus of this work is the dynamics of a spin-$0$ massless scalar field. The corresponding wave equation admits separation of variables by exploiting the Killing symmetries of the spacetime together with the generalized spherical harmonics on the three-sphere, $S^3$. Unlike the four-dimensional Schwarzschild case, the resulting radial equation is a Reduced Confluent Heun equation. For this equation, we developed an original extension of the standard MST formalism, also  providing several nontrivial consistency checks of the approach through the computation of the so-called renormalized angular momentum parameter, $\nu$, which can also be obtained independently using the quantum Seiberg-Witten formalism.

Beyond the complete construction of this formalism (centered on spin-$0$ fields but easily generalizable to other spin fields) which constitutes  an original contribution of the present work, we studied its application to the computation of the energy flux emitted by particles moving along circular orbits, an observable of particular interest in view of future gravitational wave measurements that may probe deviations from standard general relativity. We derived the Post-Newtonian expansion of the energy loss up to fractional 2.5PN order, thereby providing useful benchmarks for future higher-order analytical calculations and numerical investigations.

The fact that the MST formalism, originally developed for the Confluent Heun equation, can be successfully generalized to the Reduced Confluent Heun equation represents an important conceptual advance (maybe inferable but never explicitly shown before). We expect that this result may pave the way for further generalizations, for example to the Doubly Confluent Heun equation arising in the study of scalar wave propagation in the four-dimensional extremal Reissner--Nordstr\"om black hole. We leave this problem to future works.

We also expect that this framework will help shedding further light on the Couch--Torrence symmetry of the four-dimensional extremal Reissner--Nordstr\"om spacetime \cite{Bianchi:2021yqs,Bianchi:2022wku,Akhond:2026suo}, whose implications are still not fully understood and deserve further investigation.

Finally, we have included in Appendices useful results concerning the general $d$ dimensional Schwarzschild-Tangherlini case and the spherical harmonics on $S_3$.

\appendix

\section{Schwarzschild Tangherlini in $d$ dimensions: brief reminders}

The general Schwarzschild-Tangherlini metric in $d\ge 4$ dimensions reads
\begin{equation}
ds^2
=
-f(r)\,dt^2
+\frac{dr^2}{f(r)}
+r^2 d\Omega_{d-2}^2,
\end{equation}
where $d\Omega_{d-2}^2$ is the metric of the unit sphere in $d-2$ dimensions and
\begin{equation}
f(r)=1-\left(\frac{r_h}{r}\right)^{d-3}\,.
\end{equation}
A widely adopted notation (also used here) is $r_h=M$; the latter quantity $M$ is related to the ADM mass ${\mathcal M}$ of the spacetime
via the relation
\beq
r_h^{d-3}=\frac{16\pi G_d {\mathcal M}}{(d-2)\Omega_{d-2}}\,,
\eeq
where dimensions are such that $[G_d]=[L^{d-1} M^{-1} T^{-2}]$, and $\Omega_{d-2}$ is the volume of the unit sphere in dimensions $d-2$ 
\beq
\Omega_{d-2}=\frac{2\pi^{\frac{d-1}{2}}}{\Gamma \left( \frac{d-1}{2}\right)}\,.
\eeq
In the case $d=4$ one has $r_h \sim  \frac{G_4 {\mathcal M}}{c^2}$,  while  in $d=5$ one has $r_h^2\sim \frac{G_5 {\mathcal M}}{c^2}$, with $G_5 \sim L G_4$. Consequently, in the former case the mass will scale as $1/c^2$ whereas in the latter case (the one studied here) the mass  scales as $1/c$, i.e., an essential information when performing PN expansions in this background. 

A massless scalar field $\Phi$ satisfies the Klein-Gordon equation
\beq
\Box \Phi = 0\,.
\eeq
Due to the spacetime symmetries let us look for solutions of the type
\beq
\Phi(t,r,\Omega)=e^{-i\omega t}Y_{\ell}(\Omega)R(r),
\eeq
where $\Omega$ denotes all angular variables and the spherical harmonics $S^{d-2}$ are such that
\beq
\nabla^2_{S^{d-2}}Y_\ell=-\ell(\ell+d-3)\,Y_\ell\,.
\eeq
The resulting radial equation is given by
\begin{equation}
\label{rad_eq}
\frac{1}{r^{d-2}}
\frac{d}{dr}\left(r^{d-2}f(r)\frac{dR}{dr}\right)
+\left[\frac{\omega^2}{f(r)}-\frac{\ell(\ell+d-3)}{r^2}\right]R=0\,.
\end{equation}
Introducing the tortoise coordinate $r_*$,
\begin{equation}
\frac{dr_*}{dr}=\frac{1}{f(r)},
\end{equation}
and rescaling the radial function as  
\begin{equation}
R(r)=r^{-\frac{d-2}{2}}\psi(r),
\end{equation}
one obtains the following Schr\"odinger-like equation 
\begin{equation}
\label{rad_resc}
\frac{d^2\psi}{dr_*^2}+\left[\omega^2-V_\ell(r)\right]\psi=0\,,
\end{equation}
with effective potential   given by
\bea
V_\ell(r)&=&f(r) \left[\frac{\ell(\ell+d-3)}{r^2}+\frac{(d-2)(d-4)}{4r^2}f(r)\right.\nonumber\\
&+&\left. \frac{d-2}{2r}f'(r) \right]\,,
\eea
with 
\begin{equation}
f'(r)=(d-3)\frac{r_h^{\,d-3}}{r^{d-2}}\,.
\end{equation}
In $d=4$ one recovers the Schwarzschild scalar potential
\begin{equation}
V_\ell(r)=\left(1-\frac{2M}{r}\right)
\left[\frac{\ell(\ell+1)}{r^2}+\frac{2M}{r^3}
\right]\,.
\end{equation}

Let us add some useful considerations. The radial equation \eqref{rad_resc} in $d$ dimensions   is of Fuchsian type for any $d$. However, for $d>5$ it admits complex Fuchsian points which correspond to Cauchy horizons. Passing to the coordinate $(r/M)^{d-3}=z$, namely the one which for $d=4,5$ allows to map the equation into the corresponding Seiberg-Witten curve, the transformed equation in the variable $z$ is no more Fuchsian. 
For this reason we have chosen to discuss here the case $d=5$ and not the generic $d$ dimensional case, which indeed requires a separated treatment. 

Noticeably, in the general $d$ dimensions case several interesting quantities, e.g. the radial action for null geodesics, can be computed and studied.
 
\subsection{Radial action}

To this end, let start from Eq. \eqref{rad_eq} where 
the redshift factor reads
\beq
f(r)=1-\left(\frac{M}{r}\right)^{d-3}\,.
\eeq
Assuming
\beq
R(r)=\frac{r^{1-\frac{d}{2}}}{\sqrt{M^2-M^dr^{3-d}}}\psi(r)
\eeq
one gets the following normal form
\bea
\psi''(r)&+&Q_{\rm W,d}\psi(r)=0\,,
\eea
where
\bea
Q_{\rm W,d}&=&\frac{1}{4 r^2 \left(r^3 M^d-M^3
   r^d\right)^2}\Big[4M^6r^{2(1+d)}\omega^2\nn\\
   &+&2 \left(d^2+2 d (\ell-3)+2 (\ell-3)
   \ell+8\right) M^{d+3} r^{d+3}\nn\\
   &-&M^6
   (d{+}2 \ell{-}4) (d{+}2 \ell{-}2) r^{2
   d}{+}r^6 M^{2 d}\Big]\,.
\eea
Let us consider the \lq\lq eikonal limit" of $Q_{\rm W,d}$ replacing
\beq
\ell=\frac{J}{\hbar}=\frac{b E}{\hbar},\quad \omega=\frac{E}{\hbar}
\eeq
and expanding for small $\hbar$. The leading-order term (leading to the radial action for null  geodesics) reads
\beq
Q_{\rm geo,d}=\frac{r^{d-2}(M^6r^{d+2}+b^2(M^{d+3}r^3-M^6r^d))}{(M^dr^3-M^3r^d)^2}\,.
\eeq
One can identify (and evaluate in a large $b$ expansion limit)  two turning points: an internal 
turning point given by
\bea
r_-&=&M+\frac{M^3}{b^2 (d-3)}+\frac{(d+2) M^5}{2 b^4 (d-3)^2}\nn\\
&+&\frac{-\frac{2 M^7}{3-d}+\frac{21
   M^7}{(3-d)^2}-\frac{49
   M^7}{(3-d)^3}}{6 b^6}\nn\\
   &+&\frac{-\frac{2 M^9}{3-d}+\frac{33
   M^9}{(3-d)^2}-\frac{162
   M^9}{(3-d)^3}+\frac{243
   M^9}{(3-d)^4}}{8 b^8}\nn\\
   &+&O\left(b^{-10}\right)\,,
\eea
and an external one given by
\bea
r_+&=&b-\frac{1}{2} b^{4-d} M^{d-3}-\frac{1}{8} (2 d-5) b^{7-2 d}
   M^{2 d-6}\nn\\
   &-&\frac{1}{16} (d-2) (3 d-8)
   b^{10-3 d} M^{3 d-9}\nn\\
   &-&\frac{1}{384} (4 d-11) (4 d-9)
   (4 d-7) b^{13-4 d} M^{4 d-12}\nn\\
   &+&O\left(b^{16-5d}\right)\,.
\eea
The radial action for  null  geodesics turns out to be defined as
\beq
\label{radial_d_action}
I_r^{\rm (d)}=\int_\infty^{r_+} \sqrt{Q_{\rm geo,d}(r)}dr\,,
\eeq
and its  regular part is obtained by considering only the leading order of the turning point $r_+\sim b$.

Introducing the notation
\beq
u=\frac{b}{r},\quad \hat{b}=\frac{b}{M}\,,
\eeq
Eq. \eqref{radial_d_action} assumes the compact form
\beq
I_r^{\rm (d)}=\hat{b}M\int_0^1\frac{\sqrt{1-u^2}}{u^2}\frac{\sqrt{1+\frac{u^{d-1}}{\hat{b}^{d-3}(1-u^2)}}}{(1-\frac{u^{d-3}}{\hat{b}^{d-3}})}\,,
\eeq
which can be integrated exactly in terms of Fox-Wright ${}_p\Psi_q$ functions \cite{Miller:1995}
\beq
I_r^{(d)}=
\frac{\hat{b} M\sqrt{\pi}}{2(d-3)}
\,{}_2\Psi_2\!\left[
\begin{array}{c}
\left(-\dfrac{1}{d-3},\,1\right),
\left(\dfrac12,\,\dfrac{d-1}{2}\right)\\[2mm]
\left(1-\dfrac{1}{d-3},\,1\right),
\left(1,\,\dfrac{d-3}{2}\right)
\end{array}
;\,
\hat{b}^{\,3-d}
\right]\,,
\eeq
where
\bea
\label{fox_wright}
{}_p\Psi_q(x)=\sum_{n=0}^\infty \frac{\prod_{i=1}^p \Gamma(a_i+A_i n) z^n}{\prod_{j=1}^q \Gamma(b_j+B_j n) n! }\,,
\eea
where $a_i,b_j \in {\mathbb C}$, $A_i,B_j>0$ (in general). 
The parameters $A_i$ and $B_j$ are called \lq\lq step parameters" because they determine how the arguments of the gamma functions increase with $n$.

An equivalent representation of the Fox-Wright function ${}_p\Psi_q$ is the following 
\begin{widetext}
\bea
{}_p\Psi_q\Bigg[\begin{array}{c}
(a_1,
A_1),\dots,(a_p,A_p)\\[2mm]
(b_1,B_1),\dots,(b_q,B_q)
\end{array}
;z\Bigg]=H^{1,p}_{p,q+1}\Bigg[
\begin{array}{c}
(1-a_1,
A_1),\dots,(1-a_p,A_p)\\[2mm]
(0,1),(1-b_1,B_1),\dots,(1-b_q,B_q)
\end{array}
;-z\Bigg]\,.
\eea
[The function $H$ is also implemented in the Wolfram Mathematica language as FoxH \cite{FoxMath}.]

In terms of such FoxH the radial action for null geodesics becomes
\beq
I_r^{(d)}=\frac{\hat{b}M\sqrt{\pi}}{2(d-3)}H^{1,2}_{2,3}\Bigg[
\begin{array}{c}
\left(1+\frac{1}{d-3},1\right),\left(\frac{1}{2},\frac{d-1}{2}\right)\\[2mm]
(0,1),\left(\frac{1}{d-3},1\right),\left(0,\frac{d-3}{2}\right)
\end{array};-z\Bigg]\,.
\eeq
In $d=4$ the Fox-Wright function collaples to generalised hypergeometric functions
\bea
I_r^{d=4}&=&-\frac{\hat{b}M\pi}{2}{}_3F_2\Big[-\frac{1}{2},\frac{1}{6},\frac{5}{6};\frac{1}{2},1|\frac{27}{4\hat{b}^2}\Big]+\frac{4M}{3\hat{b}}{}_4F_3\Big[1,1,\frac{5}{3},\frac{7}{3};2,\frac{5}{2},\frac{5}{2}|\frac{27}{4\hat{b}^2}\Big]\nn\\
&=&\frac{\hat{b}M}{2}\Big[\pi\left({-}1{+}\frac{15}{16 \hat{b}^2}{+}\frac{1155}{1024 \hat{b}^4}{+}\frac{51051}{16384 \hat{b}^6}{+}\frac{47805615}{4194304
   \hat{b}^8}{+}\frac{3234846615}{67108864
   \hat{b}^{10}}\right){+}\frac{8}{3 \hat{b}^3}{+}\frac{28}{5 \hat{b}^5}{+}\frac{128}{7 \hat{b}^7}{+}\frac{4576}{63 \hat{b}^9}{+}\frac{53248}{165 \hat{b}^{11}}\Big]\nn\\
   &+&O\left(\hat{b}^{-12}\right)\,.
\eea
These results are in agreement with Refs. \cite{Ivanov:2025ozg,Bini:2025ltr,Bini:2025bll} where we show that the scattering radial action constitutes the eikonal leading order of the renormalized angular momentum $\nu$.

In $d=5$ we have
\bea
I_r^{d=5}&=&-\frac{\hat{b}M\pi}{2}{}_3F_2\Big[-\frac{1}{2},\frac{1}{4},\frac{3}{4};\frac{1}{2},1|\frac{4}{\hat{b}^2}\Big]\nn\\
&=&{-}\frac{\pi M \hat{b}}{2}\left(1{-}\frac{3}{4 \hat{b}^2}{-}\frac{35}{64 \hat{b}^4}{-}\frac{231}{256 \hat{b}^6}{-}\frac{32175}{16384 \hat{b}^8}{-}\frac{323323}{65536 \hat{b}^{10}}{-}\frac{14196819}{1048576
   \hat{b}^{12}}{-}\frac{165480975}{4194304
   \hat{b}^{14}}{-}\frac{128931743655}{1073741824
   \hat{b}^{16}}\right)\nn\\
   &+&O\left(\hat{b}^{-16}\right)\,.
\eea
A derivative with respect to $\hat{b}$ reproduce the scattering angle, whose expression matches the massless limit of \eqref{ang5d}.
\end{widetext}

\subsection{Relation with the Generalized Hypergeometric Function}

When $A_i=B_j=1,$
the Fox-Wright function reduces to
\beq
{}_p\Psi_q
=
\frac{\prod_{i=1}^{p}\Gamma(a_i)}
{\prod_{j=1}^{q}\Gamma(b_j)}
\,{}_pF_q(a_1,\ldots,a_p;
b_1,\ldots,b_q;z).
\eeq
Hence, the Fox-Wright function is a genuine extension of the generalized hypergeometric function.

\subsection{Convergence}

Let us introduce the notation
\beq
{\mathcal A}_p=\sum_{i=1}^{p}A_i\,,\quad {\mathcal B}_q=\sum_{j=1}^{q}B_j\,,
\eeq
and let us denote
\beq
{\mathcal D}=1+{\mathcal B}_q-{\mathcal A}_p\,.
\eeq
The above series \eqref{fox_wright} converges for every complex value of $z$ whenever
${\mathcal D}>0$. If ${\mathcal D}=0$ the series has a finite radius of convergence depending on the parameters while
when ${\mathcal D}<0$ the series generally diverges for $z\neq0$, although it may still represent an asymptotic expansion.

\subsection{Special Cases}

{\it The Wright Function}\\

The Wright function is obtained as
\beq
{}_0\Psi_1\!\left[
\begin{matrix}
-\\
(\beta,\alpha)
\end{matrix};z
\right]
=
\sum_{k=0}^{\infty}
\frac{z^k}{k!\,\Gamma(\alpha k+\beta)}.
\eeq

{\it The Mittag-Leffler Function}\\

The two-parameter Mittag-Leffler function can be written as
\beq
E_{\alpha,\beta}(z)
=
\sum_{k=0}^{\infty}
\frac{z^k}{\Gamma(\alpha k+\beta)}
=
{}_1\Psi_1
\!\left[
\begin{matrix}
(1,1)\\
(\beta,\alpha)
\end{matrix};z
\right].
\eeq

{\it The Generalized Hypergeometric Function}\\

When all step parameters are equal to one, one recovers the generalized hypergeometric function:
\beq
{}_pF_q
\subset
{}_p\Psi_q.
\eeq

\subsection{Relation with the Fox $H$-Function}

The Fox-Wright function can often be represented as a particular case of the Fox $H$-function,
\beq
{}_p\Psi_q(z)=H_{p,q+1}^{1,p}\!\left(-z\;|\;\cdots\right),
\eeq
which places it within the hierarchy of generalized special functions together with the Meijer $G$-function and the generalized hypergeometric function.

The Fox-Wright function therefore provides a unifying framework for a large class of classical and fractional-calculus special functions and plays an important role in modern analysis.

Interestingly, it is the first time that the ${}_p\Psi_q$ function appears in this and similar contexts.
We have just recalled how the Fox-Wright ${}_p\Psi_q$ function is a special case of Fox's $H$-function and a generalization of the generalized hypergeometric function. As shown above, 
the ${}_p\Psi_q$  function reduce to a single generalized hypergeometric function when certain parameters are integers,  and to a finite sum of generalized hypergeometric functions when these parameters are rational numbers. In general, ${}_p\Psi_q$  function it is also connected with the Meijer's G-function.

In the literature \cite{Miller:1995} the Fox--Wright function appears in numerous areas of mathematics and mathematical physics, including fractional differential equations, fractional calculus, probability distributions with heavy tails, anomalous diffusion, integral transforms, generalized kinetic equations, statistical mechanics and quantum physics.

\subsection{PN-type solution to the radial equation \eqref{rad_eq}}

We will not aim at discussing here in detail the generic $d$ solution in comparison with the corresponding one in $d=4$, since it deserves a specific treatment as well as explicit examples in a context of dimensional regularization when $d=4+\epsilon$. Dedicated studies on this topic are still in progress and will be discussed in future works~\cite{davide}. 

Let us just mention, in passing, the following property concerning the PN-type solution (we will refer only to the in-solution since the up-solution can be obtained from the in-solution via the replacing $\ell\to -\ell-d+3$).

Starting from $R_{\rm in}^{(d=4+\epsilon)}(r)$ in the limit $\epsilon \to 0$ one gets
\bea
R_{\rm in}^{(d=4+\epsilon)}(r)=\frac{R_{-1}}{\epsilon} +R_0+O(\epsilon)\,,
\eea
where $R_0=R_{\rm in}^{(d=4)}(r)$ and
\beq
R_{-1}=-\frac{\nu_{\ell}-\ell}{8} R_0\,,
\eeq
where $\nu_{\ell}$ denotes the 4d renormalized angular momentum variable (familiar from previous works in the Schwarzschild case in the context of scalar self force).
This sort of \lq\lq dimensional regularization" is expected to play a role when other types of regularizations were not effective.
 
\section{Geometrical characterizations of $S_3$: brief reminders}

In this appendix we will recall some geometrical properties of the 3-sphere $S_3$.

\subsection{Metric and curvature}

Let us write the metric of $S_3$ in the form (see Eq. \eqref{met_split})
\beq
\label{3-sp}
ds^2=d\theta^2 +\sin^2\theta d\phi^2 +\cos^2\theta d\psi^2\,,
\eeq
where the coordinates are $x^a=(\theta,\phi,\psi)$ vary in the ranges $\theta \in [0,\frac{\pi}{2})$, $\phi,\psi \in [0,2\pi)$.
The spacetime \eqref{3-sp}  is an Einstein manifold because of the property
\beq
R_{ab}-2g_{ab}=0\,,\quad R=6\,,
\eeq
also implying $R_{ab;c}=0$.
The independent Riemann tensor components are
\bea
&& R_{t\phi t\phi}= \sin^2 \theta \,,\quad R_{t\psi t\psi}= \cos^2 \theta,\nonumber\\
&& R_{\phi\psi\phi\psi}= \sin^2 \theta \cos^2 \theta \,.
\eea
We recall that in $d=3$ for an Einstein metric  the Weyl tensor vanishes identically.

Furthermore, because of the Killing vectors $\partial_\phi$ and $\partial_\psi$ there exist two conserved quantities along geodesics (parametrized by the affine parameter $\lambda$)
\beq
\frac{d\phi}{d\lambda} = \frac{L_\phi}{\sin^2 \theta}\,,\qquad
\frac{d\psi}{d\lambda}  = \frac{L_\psi}{\cos^2\theta}\,.
\eeq
The remaining equation
\beq
\frac{d^2\theta}{d\lambda^2}-\frac{L_\phi^2 \cos\theta}{\sin^3 \theta}+\frac{L_\psi^2 \sin\theta}{\cos^3 \theta}=0
\eeq
can be solved by quadratures. In fact, multiplying both sides of this equation  by $2 \frac{d\theta}{d\lambda}$, one finds
\beq
\left(\frac{d\theta}{d\lambda}\right)^2 +  \frac{L_\phi^2}{\sin^2\theta}+ \frac{L_\psi^2}{\cos^2\theta}=K^2\ge 0\,,
\eeq
with $K^2$ a (non-negative) constant.

Let us denote
\beq
{\mathcal L}_\pm =\frac{L_\psi\pm L_\phi}{K}\,.
\eeq
The solution of the previous equation is
\bea
\theta(\lambda)&=&\frac{1}{2}{\rm arccos}\Big[A_{s^0}+A_{s^1}\sin\left(2K\lambda\right)\Big]\,,
\eea
with
\bea
A_{s^0}&=&{\mathcal L}_+{\mathcal L}_- \,,\nn\\ 
A_{s^1}&=&K^2 \sqrt{(1-{\mathcal L}_+)(1-{\mathcal L}_-)}\,.
\eea
The behavior of the $\theta$-motion follows by introducing the effective potential
\beq
V_{\rm eff}=\frac{L_\phi^2}{\sin^2\theta}+ \frac{L_\psi^2}{\cos^2\theta}\,,
\eeq
which admits a minimum in $\theta_{\rm min}={\rm arctan}\left(\sqrt{\frac{L_\phi}{L_\psi}} \right)$, with
\beq
\sin \theta_{\rm min}=\sqrt{\frac{L_\phi}{L_\phi+L_\psi}}, \quad \cos\theta_{\rm min}=\sqrt{\frac{L_\psi}{L_\phi+L_\psi}}\,,
\eeq
and
\beq
V_{\rm eff}|_{\rm min}=(L_\phi+L_\psi)^2\,.
\eeq
Clearly, if $K^2>V_{\rm eff}|_{\rm min}$ the $\theta$ motion corresponds to  oscillations between two (symmetric) $\theta$ values (easily determined analytically but corresponding to rather involved expressions); if $K^2=V_{\rm eff}|_{\rm min}$ $\theta$ is  constant, while for $K^2<V_{\rm eff}|_{\rm min}$ the $\theta$ motion is impossible.\\

\subsection{Wigner matrices on $S^3 \simeq SU(2)$}

Since $S^3$ is diffeomorphic to the compact group   $SU(2)$, the Wigner matrices
$D^{\,\ell}_{m_1m_2}(g)$, with $g\in SU(2)$, form an orthogonal basis of the space $L^2(SU(2))$.
Parametrizing  $g$ by the Euler angles, say  
$(\phi,\theta,\psi)$, and introducing the generators of the rotations $J_i$
\beq
g=e^{-i\phi J_z}e^{-i\theta J_y}e^{-i\psi J_z},
\eeq
the matrix elements of the irreducible, unitary  representations of $SU(2)$
are of the type
\beq
D^{\,\ell}_{m_1m_2}(\phi,\theta,\psi)=
e^{-im_1\phi}\,
d^{\,\ell}_{m_1m_2}(\theta)\,
e^{-im_2\psi},
\eeq
where $d^{\,\ell}_{m_1m_2}(\theta)$ are the so called little Wigner matrices.

\subsubsection{Ortogonality}

The functions $D^{\,\ell}_{m_1m_2}$ satisfy the orthogonality relation
\beq
\int_{SU(2)}
D^{\,\ell}_{m_1m_2}(g)\,
D^{*\,\ell'}_{n_1n_2}(g) 
\, dg
=
\frac{1}{2\ell+1}
\delta_{\ell \ell'}
\delta_{m_1n_1}
\delta_{m_2n_2},
\eeq
where $dg$ denotes the Haar measure  normalized on  $SU(2)$.
Using Euler's coordinates,
\beq
dg=
\frac{1}{8\pi^2}
\sin\theta\,
d\phi\,d\theta\,d\psi,
\eeq
and hence
\begin{widetext}
\beq
\int_0^{2\pi} d\phi
\int_0^\pi d\theta
\int_0^{2\pi} d\psi\,
\sin\theta\,
D^{\,\ell}_{m_1m_2}(\phi,\theta,\psi)
D^{\,\ell' *}_{n_1n_2}(\phi,\theta,\psi) 
=
\frac{8\pi^2}{2\ell+1}
\delta_{\ell \ell'}
\delta_{m_1n_1}
\delta_{m_2n_2}.
\eeq
\end{widetext}
This relation represents the analogous on $S_3$ of the orthogonality relation of the spherical harmonics in $S_2$
\beq
\int_{S^2}
Y_{\ell m}(\Omega)
Y^*_{\ell'm'}(\Omega)
\, d\Omega
=
\delta_{\ell \ell'}\delta_{mm'}.
\eeq

\subsubsection{Expression in terms of Jacobi polynomials}

The functions $d^{\,\ell}_{m_1m_2}(\theta)$ can be expressed in terms of Jacobi polynomials
\bea
d^{\,\ell}_{m_1m_2}(\theta)
&=&
N_{\ell m_1m_2}\,
\left(\sin\frac{\theta}{2}\right)^{m_1-m_2}
\left(\cos\frac{\theta}{2}\right)^{m_1+m_2}\times \nonumber\\
&&
P_{\,\ell-m_1}^{(m_1-m_2,m_1+m_2)}
(\cos\theta),
\eea
where $N_{\ell m_1m_2}$ are proper normalization coefficients.
Since
\bea
&&P_n^{(\alpha,\beta)}(x)
=
\frac{(\alpha+1)_n}{n!}\times \,\nonumber\\
&&
{}_2F_1
\!\left(
-n,\,
n+\alpha+\beta+1;\,
\alpha+1;\,
\frac{1-x}{2}
\right),
\eea
one finds that the Wigner matrices can be expressed in terms of Gauss hypergeometric functions   ${}_2F_1$.

\subsubsection{Harmonic basis on $S^3$}

Wigner's matrices are the natural generalization of the standard  spherical harmonics in $S_3$
\bea
&& S^1:\quad e^{im\phi},\nonumber\\
&& S^2:\quad Y_{\ell m}(\theta,\phi)\,,\nonumber\\
&& S^3:\quad D^{\,\ell}_{m_1m_2}(\phi,\theta,\psi).
\eea
They diagonalize the laplacian of  $S^3$:
\beq
\Delta_{S^3}
D^{\,\ell}_{m_1m_2}
=
-4\,\ell(\ell+1)\,
D^{\,\ell}_{m_1m_2},
\eeq
where the numerical factor depends on the adopted convention for the radius of the sphere and for the normalization of the Laplacian.
Therefore,  the functions $D^{\,\ell}_{m_1m_2}$ form a complete orthonormal basis of  $L^2(S^3)$, in agreement with Peter-Weyl's theorem  for the (compact) group 
$SU(2)$.

\section{Geometric characterization of the $t-r$ part of the metric in $d=5$}

Let us study the geometrical properties of the $2$-metric
\bea
\label{t_r_metric}
{}_{(2)} ds_{(t,r)}^2&=& g_{AB}dx^A dx^B\nonumber\\
&=& -\left(1-\frac{M^2}{r^2} \right)dt^2+\frac{dr^2}{\left(1-\frac{M^2}{r^2} \right)}\,,
\eea
referred to coordinates $x^A=(t,r)$, see Eq. \eqref{met_split}. The metric is not Ricci flat but
\beq
R_{AB}-\frac{3M^2}{r^4}g_{AB}=0\,,
\eeq
namely it is again an Einstein metric (but now the factor in front of the metric is not a constant).
The associated scalar curvature is given by
\beq
R=\frac{6M^2}{r^4}\,,
\eeq
and the only nonvanishing component of the Riemann tensor is
\beq
R_{trtr}= -\frac{3M^2}{r^4}=-\frac12 R\,.
\eeq
As a minimum of complementary information to the angular part of the spacetime geometry, we recall that timelike geodesics correspond to the unit timelike vector
\beq
U=\frac{E}{1-\frac{M^2}{r^2}}\partial_t \pm \left(E^2-1+\frac{M^2}{r^2}  \right)^{1/2}\partial_r\,,
\eeq
where $E$ is the conserved energy per unit of the probe  mass.
Orbits at fixed $r=r_0$ have
\beq
\label{r0_def}
r_0=\frac{M}{\sqrt{1-E^2}}
\eeq 
whereas the general solution of the radial equation $\frac{dr}{d\tau}=U^r$ reads
\beq
r(\tau) =\sqrt{(E^2-1)\tau^2 +r_0^2}\,,
\eeq
with $r_0=r(0)$ given in Eq. \eqref{r0_def},
leading to open/closed orbits according the value (positive/negative) of $E^2-1$.

\section*{Acknowledgments}

We thank  
G.~Aminov, A.~Geralico, A.~Grassi, Y.~Hatsuda, T.~Manton for useful comments and discussions. 
V.~F. acknowledges hospitality at Istituto per le Applicazioni del Calcolo \lq\lq M. Picone" CNR, Rome at various stages during the development of the present project.
D.~B. and G.~D.~R. acknowledge membership to the Italian Gruppo Nazionale per la Fisica Matematica (GNFM) of the Istituto Nazionale di Alta Matematica
(INDAM). 
\\

\section*{Data availability} 
The data that support the findings of this article are openly available \cite{dataval}.


\begin{thebibliography}{99}

\bibitem{Tangherlini:1963bw}
F.~R.~Tangherlini,
``Schwarzschild field in n dimensions and the dimensionality of space problem,''
Nuovo Cim. \textbf{27}, 636-651 (1963)
doi:10.1007/BF02784569

\bibitem{Gregory:1993vy}
R.~Gregory and R.~Laflamme,
``Black strings and p-branes are unstable,''
Phys. Rev. Lett. \textbf{70}, 2837-2840 (1993)
doi:10.1103/PhysRevLett.70.2837
[arXiv:hep-th/9301052 [hep-th]].

\bibitem{HOROWITZ1991197}
G.~T.~Horowitz and A.~Strominger,
``Black strings and p-branes,"
Nuclear Physics B \textbf{360-1}, 197-209 (1991)
doi:10.1016/0550-3213(91)90440-9.

\bibitem{Myers:1986un}
R.~C.~Myers and M.~J.~Perry,
``Black Holes in Higher Dimensional Space-Times,''
Annals Phys. \textbf{172}, 304 (1986)
doi:10.1016/0003-4916(86)90186-7

\bibitem{Cardoso:2005gj}
V.~Cardoso, O.~J.~C.~Dias, J.~L.~Hovdebo and R.~C.~Myers,
``Instability of non-supersymmetric smooth geometries,''
Phys. Rev. D \textbf{73}, 064031 (2006)
doi:10.1103/PhysRevD.73.064031
[arXiv:hep-th/0512277 [hep-th]].

\bibitem{Bianchi:2019lmi}
M.~Bianchi, M.~Casolino and G.~Rizzo,
Nucl. Phys. B \textbf{954}, 115010 (2020)
doi:10.1016/j.nuclphysb.2020.115010
[arXiv:1904.01097 [hep-th]].

\bibitem{Bianchi:2022qph}
M.~Bianchi and G.~Di Russo,
``2-charge circular fuzz-balls and their perturbations,''
JHEP \textbf{08}, 217 (2023)
doi:10.1007/JHEP08(2023)217
[arXiv:2212.07504 [hep-th]].

\bibitem{Bianchi:2023rlt}
M.~Bianchi, C.~Di Benedetto, G.~Di Russo and G.~Sudano,
``Charge instability of JMaRT geometries,''
JHEP \textbf{09}, 078 (2023)
doi:10.1007/JHEP09(2023)078
[arXiv:2305.00865 [hep-th]].

\bibitem{DiRusso:2024hmd}
G.~Di Russo, F.~Fucito and J.~F.~Morales,
``Tidal resonances for fuzzballs,''
JHEP \textbf{04}, 149 (2024)
doi:10.1007/JHEP04(2024)149
[arXiv:2402.06621 [hep-th]].

\bibitem{Bah:2020pdz}
I.~Bah and P.~Heidmann,
``Topological stars, black holes and generalized charged Weyl solutions,''
JHEP \textbf{09}, 147 (2021)
doi:10.1007/JHEP09(2021)147
[arXiv:2012.13407 [hep-th]].

\bibitem{Bianchi:2023sfs}
M.~Bianchi, G.~Di Russo, A.~Grillo, J.~F.~Morales and G.~Sudano,
``On the stability and deformability of top stars,''
JHEP \textbf{12}, 121 (2023)
doi:10.1007/JHEP12(2023)121
[arXiv:2305.15105 [gr-qc]].

\bibitem{Heidmann:2023ojf}
P.~Heidmann, N.~Speeney, E.~Berti and I.~Bah,
``Cavity effect in the quasinormal mode spectrum of topological stars,''
Phys. Rev. D \textbf{108}, no.2, 024021 (2023)
doi:10.1103/PhysRevD.108.024021
[arXiv:2305.14412 [gr-qc]].

\bibitem{Cipriani:2024ygw}
A.~Cipriani, C.~Di Benedetto, G.~Di Russo, A.~Grillo and G.~Sudano,
``Charge (in)stability and superradiance of Topological Stars,''
JHEP \textbf{07}, 143 (2024)
doi:10.1007/JHEP07(2024)143
[arXiv:2405.06566 [hep-th]].

\bibitem{Bianchi:2024vmi}
M.~Bianchi, D.~Bini and G.~Di Russo,
``Scalar perturbations of topological-star spacetimes,''
Phys. Rev. D \textbf{110}, no.8, 084077 (2024)
doi:10.1103/PhysRevD.110.084077
[arXiv:2407.10868 [gr-qc]].

\bibitem{Bianchi:2024rod}
M.~Bianchi, D.~Bini and G.~Di Russo,
``Scalar waves in a topological star spacetime: Self-force and radiative losses,''
Phys. Rev. D \textbf{111}, no.4, 044017 (2025)
doi:10.1103/PhysRevD.111.044017
[arXiv:2411.19612 [gr-qc]].

\bibitem{DiRusso:2025lip}
G.~Di Russo, M.~Bianchi and D.~Bini,
``Scalar waves from unbound orbits in a topological star spacetime: PN reconstruction of the field and radiation losses in a self-force approach,''
Phys. Rev. D \textbf{112}, no.2, 024002 (2025)
doi:10.1103/sycf-brn1
[arXiv:2502.21040 [gr-qc]].

\bibitem{Heidmann:2025pbb}
P.~Heidmann, P.~Pani and J.~E.~Santos,
``Asymptotically flat rotating topological stars,''
JHEP \textbf{03}, 108 (2026)
doi:10.1007/JHEP03(2026)108
[arXiv:2510.05200 [hep-th]].

\bibitem{Bianchi:2025uis}
M.~Bianchi, G.~Dibitetto, J.~F.~Morales and A.~Ruip{\'e}rez,
``Rotating Topological Stars,''
JHEP \textbf{01}, 046 (2026)
doi:10.1007/JHEP01(2026)046
[arXiv:2504.12235 [hep-th]].

\bibitem{Dima:2025tjz}
A.~Dima, P.~Heidmann, M.~Melis, P.~Pani and G.~Patashuri,
``W-solitons as prototypical black hole microstates,''
Phys. Rev. D \textbf{112}, no.12, 124056 (2025)
doi:10.1103/2wcq-4xny
[arXiv:2509.18245 [gr-qc]].

\bibitem{Mano:1996mf}
S.~Mano, H.~Suzuki and E.~Takasugi,
``Analytic solutions of the Regge-Wheeler equation and the postMinkowskian expansion,''
Prog. Theor. Phys. \textbf{96}, 549-566 (1996)
doi:10.1143/PTP.96.549
[arXiv:gr-qc/9605057 [gr-qc]].

\bibitem{Mano:1996vt}
S.~Mano, H.~Suzuki and E.~Takasugi,
``Analytic solutions of the Teukolsky equation and their low frequency expansions,''
Prog. Theor. Phys. \textbf{95}, 1079-1096 (1996)
doi:10.1143/PTP.95.1079
[arXiv:gr-qc/9603020 [gr-qc]].

\bibitem{Sasaki:2003xr}
M.~Sasaki and H.~Tagoshi,
``Analytic black hole perturbation approach to gravitational radiation,''
Living Rev. Rel. \textbf{6}, 6 (2003)
doi:10.12942/lrr-2003-6
[arXiv:gr-qc/0306120 [gr-qc]].


\bibitem{Chandrasekhar:1985kt}
S.~Chandrasekhar,
``The mathematical theory of black holes,''
Oxford Classic Texts in the Physical Sciences, Ed. revis. 1998,
Oxford University Press, USA

\bibitem{Parnachev:2020zbr}
A.~Parnachev and K.~Sen,
``Notes on AdS-Schwarzschild eikonal phase,''
JHEP \textbf{03}, 289 (2021)
doi:10.1007/JHEP03(2021)289
[arXiv:2011.06920 [hep-th]].

\bibitem{Ivanov:2025ozg}
M.~M.~Ivanov, Y.~Z.~Li, J.~Parra-Martinez and Z.~Zhou,
``Resummation of Universal Tails in Gravitational Waveforms,''
Phys. Rev. Lett. \textbf{135}, no.14, 141401 (2025)
doi:10.1103/jzd1-qzkt
[arXiv:2504.07862 [hep-th]].

\bibitem{Bini:2025ltr}
D.~Bini and G.~Di Russo,
``Topological stars and scalar wave equation: Exact resummation of the renormalized angular momentum in the eikonal limit,''
Phys. Rev. D \textbf{112}, no.6, 064008 (2025)
doi:10.1103/dw5y-4pv8
[arXiv:2506.14442 [gr-qc]].

\bibitem{Bini:2025bll}
D.~Bini, G.~Di Russo and A.~Geralico,
``Kerr spacetime and scalar wave equation: Exact resummation of the renormalized angular momentum in the eikonal limit,''
Phys. Rev. D \textbf{112}, no.6, 064077 (2025)
doi:10.1103/mzqw-wbvf
[arXiv:2508.12046 [gr-qc]].


\bibitem{Damour:1988mr}
T.~Damour and G.~Schaefer,
``Higher Order Relativistic Periastron Advances and Binary Pulsars,''
Nuovo Cim. B \textbf{101}, 127 (1988)
doi:10.1007/BF02828697


\bibitem{Bini:2025qyn}
D.~Bini and G.~Di Russo,
``Characterizing geodesic deviations in a topological star spacetime: Massive, charged, spinning, and stringylike objects,''
Phys. Rev. D \textbf{112}, no.2, 024021 (2025)
doi:10.1103/6xc4-wm2f
[arXiv:2505.13020 [gr-qc]].

\bibitem{Bini:2004sy}
D.~Bini, C.~Cherubini, G.~Cruciani and R.~T.~Jantzen,
``Geometric transport along circular orbits in stationary axisymmetric spacetimes,''
Int. J. Mod. Phys. D \textbf{13}, 1771-1804 (2004)
doi:10.1142/S0218271804005237
[arXiv:gr-qc/0407004 [gr-qc]].

\bibitem{Cardoso:2008bp}
V.~Cardoso, A.~S.~Miranda, E.~Berti, H.~Witek and V.~T.~Zanchin,
``Geodesic stability, Lyapunov exponents and quasinormal modes,''
Phys. Rev. D \textbf{79}, no.6, 064016 (2009)
doi:10.1103/PhysRevD.79.064016
[arXiv:0812.1806 [hep-th]].

\bibitem{Bianchi:2021mft}
M.~Bianchi, D.~Consoli, A.~Grillo and J.~F.~Morales,
``More on the SW-QNM correspondence,''
JHEP \textbf{01}, 024 (2022)
doi:10.1007/JHEP01(2022)024
[arXiv:2109.09804 [hep-th]].

\bibitem{Berti:2009wx}
E.~Berti, V.~Cardoso and P.~Pani,
``Breit-Wigner resonances and the quasinormal modes of anti-de Sitter black holes,''
Phys. Rev. D \textbf{79}, 101501 (2009)
doi:10.1103/PhysRevD.79.101501
[arXiv:0903.5311 [gr-qc]].

\bibitem{Pani:2013pma}
P.~Pani,
``Advanced Methods in Black-Hole Perturbation Theory,''
Int. J. Mod. Phys. A \textbf{28}, 1340018 (2013)
doi:10.1142/S0217751X13400186
[arXiv:1305.6759 [gr-qc]].

\bibitem{Cardoso:2014sna}
V.~Cardoso, L.~C.~B.~Crispino, C.~F.~B.~Macedo, H.~Okawa and P.~Pani,
``Light rings as observational evidence for event horizons: long-lived modes, ergoregions and nonlinear instabilities of ultracompact objects,''
Phys. Rev. D \textbf{90}, no.4, 044069 (2014)
doi:10.1103/PhysRevD.90.044069
[arXiv:1406.5510 [gr-qc]].

\bibitem{Berti:2005gp}
E.~Berti, V.~Cardoso and M.~Casals,
``Eigenvalues and eigenfunctions of spin-weighted spheroidal harmonics in four and higher dimensions,''
Phys. Rev. D \textbf{73}, 024013 (2006)
[erratum: Phys. Rev. D \textbf{73}, 109902 (2006)]
doi:10.1103/PhysRevD.73.109902
[arXiv:gr-qc/0511111 [gr-qc]].

\bibitem{Bonelli:2022ten}
G.~Bonelli, C.~Iossa, D.~Panea Lichtig and A.~Tanzini,
``Irregular Liouville Correlators and Connection Formulae for Heun Functions,''
Commun. Math. Phys. \textbf{397}, no.2, 635-727 (2023)
doi:10.1007/s00220-022-04497-5
[arXiv:2201.04491 [hep-th]].

\bibitem{Aminov:2020yma}
G.~Aminov, A.~Grassi and Y.~Hatsuda,
``Black Hole Quasinormal Modes and Seiberg{\textendash}Witten Theory,''
Annales Henri Poincare \textbf{23}, no.6, 1951-1977 (2022)
doi:10.1007/s00023-021-01137-x
[arXiv:2006.06111 [hep-th]].

\bibitem{Consoli:2022eey}
D.~Consoli, F.~Fucito, J.~F.~Morales and R.~Poghossian,
``CFT description of BH{\textquoteright}s and ECO{\textquoteright}s: QNMs, superradiance, echoes and tidal responses,''
JHEP \textbf{12}, 115 (2022)
doi:10.1007/JHEP12(2022)115
[arXiv:2206.09437 [hep-th]].

\bibitem{Aminov:2023jve}
G.~Aminov, P.~Arnaudo, G.~Bonelli, A.~Grassi and A.~Tanzini,
``Black hole perturbation theory and multiple polylogarithms,''
JHEP \textbf{11}, 059 (2023)
doi:10.1007/JHEP11(2023)059
[arXiv:2307.10141 [hep-th]].

\bibitem{Bianchi:2025ydq}
M.~Bianchi, D.~Bini and G.~Di Russo,
``Scalar self-force effects in neutral W-soliton backgrounds,''
Phys. Rev. D \textbf{113}, no.4, 044028 (2026)
doi:10.1103/rf8g-yl2d
[arXiv:2511.01402 [gr-qc]].

\bibitem{Fioravanti:2025bts}
D.~Fioravanti and M.~Rossi,
``Regular and Floquet bases for gauge and gravity theories: a non perturbative approach,''
[arXiv:2508.19960 [hep-th]].

\bibitem{Nekrasov:2009rc}
N.~A.~Nekrasov and S.~L.~Shatashvili,
``Quantization of Integrable Systems and Four Dimensional Gauge Theories,''
Contribution to: ICMP09, 265-289 (2009)
doi:10.1142/9789814304634{\_}0015
[arXiv:0908.4052 [hep-th]].

\bibitem{Poghosyan:2020zzg}
H.~Poghosyan,
``Recursion relation for instanton counting for SU(2) $ \mathcal{N} $ = 2 SYM in NS limit of $\Omega$ background,''
JHEP \textbf{05}, 088 (2021)
doi:10.1007/JHEP05(2021)088
[arXiv:2010.08498 [hep-th]].

\bibitem{Cipriani:2025ikx}
A.~Cipriani, G.~Di Russo, F.~Fucito, J.~F.~Morales, H.~Poghosyan and R.~Poghossian,
``Resumming post-Minkowskian and post-Newtonian gravitational waveform expansions,''
SciPost Phys. \textbf{19}, no.2, 057 (2025)
doi:10.21468/SciPostPhys.19.2.057
[arXiv:2501.19257 [gr-qc]].

\bibitem{Mitschi:2016fxp}
C.~Mitschi and D.~Sauzin,
``Divergent Series, Summability and Resurgence I,''
Lect. Notes Math. \textbf{2153}, pp. xxi+298,
Springer, 2016,
doi:10.1007/978-3-319-28736-2

\bibitem{Ecalle:1981}
J~\'{E}calle,
``Les fonctions r\'{e}surgentes. {T}ome {I},''
Publications Math\'{e}matiques d'Orsay 81 [Mathematical
              Publications of Orsay 81] ,
{\bf 5},
Universit\'{e} de Paris-Sud, D\'{e}partement de Math\'{e}matiques, Orsay, 247 (1981)


\bibitem{Ramis:1993}
J.~P.~Ramis 
``S\'{e}ries divergentes et th\'{e}ories asymptotiques,''
 {Bull. Soc. Math. France},
{\bf 121},74 (1993).

\bibitem{diver-ii}
M.~Loday-Richaud, 
``Divergent series, summability and resurgence. II'', 
Lect. Notes Math., \textbf{2154}, pp. xxiii+272, 
Springer, (2016) 
doi:10.1007/978-3-319-29075-1




\bibitem{Bini:2016egn}
D.~Bini, G.~Carvalho and A.~Geralico,
``Scalar field self-force effects on a particle orbiting a Reissner-Nordstr{\"o}m black hole,''
Phys. Rev. D \textbf{94}, no.12, 124028 (2016)
doi:10.1103/PhysRevD.94.124028
[arXiv:1610.02235 [gr-qc]].

\bibitem{Bianchi:2021yqs}
M.~Bianchi and G.~Di Russo,
``Turning black holes and D-branes inside out of their photon spheres,''
Phys. Rev. D \textbf{105}, no.12, 126007 (2022)
doi:10.1103/PhysRevD.105.126007
[arXiv:2110.09579 [hep-th]].

\bibitem{Bianchi:2022wku}
M.~Bianchi and G.~Di Russo,
``Turning rotating D-branes and black holes inside out their photon-halo,''
Phys. Rev. D \textbf{106}, no.8, 086009 (2022)
doi:10.1103/PhysRevD.106.086009
[arXiv:2203.14900 [hep-th]].

\bibitem{Akhond:2026suo}
M.~Akhond, M.~Bianchi, A.~Cristofaro and F.~Riccioni,
``Couch-Torrence conformal inversion, supersymmetry and conserved charges for D3-branes,''
[arXiv:2604.27815 [hep-th]].

\bibitem{Miller:1995}
A.~R.~Miller, I.~S.~Moskowitz,
``Reduction of a class of Fox-Wright psi functions for certain rational parameters,''
Computers \& Mathematics with Applications,
\textbf{30}, Issue 11, 73-82,
(1995) 
ISSN 0898-1221,
https://doi.org/10.1016/0898-1221(95)00165-U.

\bibitem{FoxMath}
Wolfram Research,
\emph{FoxH},
Wolfram Language Documentation,
\url{https://reference.wolfram.com/language/ref/FoxH.html}.

\bibitem{davide}
D.~Usseglio, personal communication.

 
\bibitem{dataval}
The supplemental material included both in the arxiv version and in the published version of the present paper contains all data that support the findings of this article.

\end{thebibliography}
\end{document}